\documentclass{cernrep}
\newcommand{\mean}[1]{\left\langle #1 \right\rangle}
\newcommand{\gsim}{\mathrel{\hbox{\rlap{\lower.55ex \hbox{$\sim$}} \kern-.3em \raise.4ex \hbox{$>$}}}}
\newcommand{\lsim}{\mathrel{\hbox{\rlap{\lower.55ex \hbox{$\sim$}} \kern-.3em \raise.4ex \hbox{$<$}}}}

\begin{document}
\title{Relativistic heavy-ion collisions\footnote{Updated version of the 
lectures given at the First Asia-Europe-Pacific School of High-Energy Physics,
Fukuoka, Japan, 14-27 October 2012. Published
as a CERN Yellow Report (CERN-2014-001) and KEK report 
(KEK-Proceedings-2013-8), K. Kawagoe and M. Mulders (eds.), 2014, p. 219.}}
\author{Rajeev S. Bhalerao
%\thanks {On leave from another institue somewhere.}
}
\institute{Department of Theoretical Physics, Tata Institute of Fundamental 
Research, Mumbai, India}
\maketitle

\begin{abstract}
The field of relativistic heavy-ion collisions is introduced to the
high-energy physics students with no prior knowledge in this area. The
emphasis is on the two most important observables, namely the azimuthal
collective flow and jet quenching, and on the role fluid dynamics
plays in the interpretation of the data. Other important observables
described briefly are constituent quark number scaling, ratios of
particle abundances, strangeness enhancement, and sequential melting
of heavy quarkonia. Comparison is made of some of the basic heavy-ion
results obtained at LHC with those obtained at RHIC. Initial findings
at LHC which seem to be in apparent conflict with the accumulated RHIC
data are highlighted.
\end{abstract}

\section{Introduction}

These are exciting times if one is working in the area of relativistic
heavy-ion collisions, with two heavy-ion colliders namely the
Relativistic Heavy-Ion Collider (RHIC) at the Brookhaven National
Laboratory and the Large Hadron Collider (LHC) at CERN in operation in
tandem. Quark-gluon plasma has been discovered at RHIC, but its
precise properties are yet to be established. With the phase diagram
of strongly interacting matter (QCD phase diagram) also being largely
unknown, these are also great times for fresh graduate students to get
into this area of research, which is going to remain very active for
the next decade at least. The field is maturing as evidenced by the
increasing number of text books that are now available
\cite{satz,flor,bart,vogt,yagi,lete,wong,cser,mull}. 
Also available are collected review
articles; see \eg \cite{Friman:2011zz,hwaw,hwa}.

This is a fascinating inter-disciplinary area of research at the
interface of particle physics and high-energy nuclear physics. It
draws heavily from QCD --- perturbative, non-perturbative, as well as 
semiclassical. It has overlaps with thermal field theory,
relativistic fluid dynamics, kinetic or transport theory, quantum
collision theory, apart from the standard statistical mechanics and
thermodynamics. Quark-Gluon Plasma (QGP) at high temperature, $T$, and
vanishing net baryon number density, $n_B$ (or equivalently the
corresponding chemical potential, $\mu_B$), is of cosmological
interest, while QGP at low $T$ and large $n_B$ is of astrophysical
interest. String theorists too have developed interest in this area
because of the black hole -- fluid dynamics connection.

\begin{figure}[htbp]
\centering\includegraphics[width=.8\linewidth]{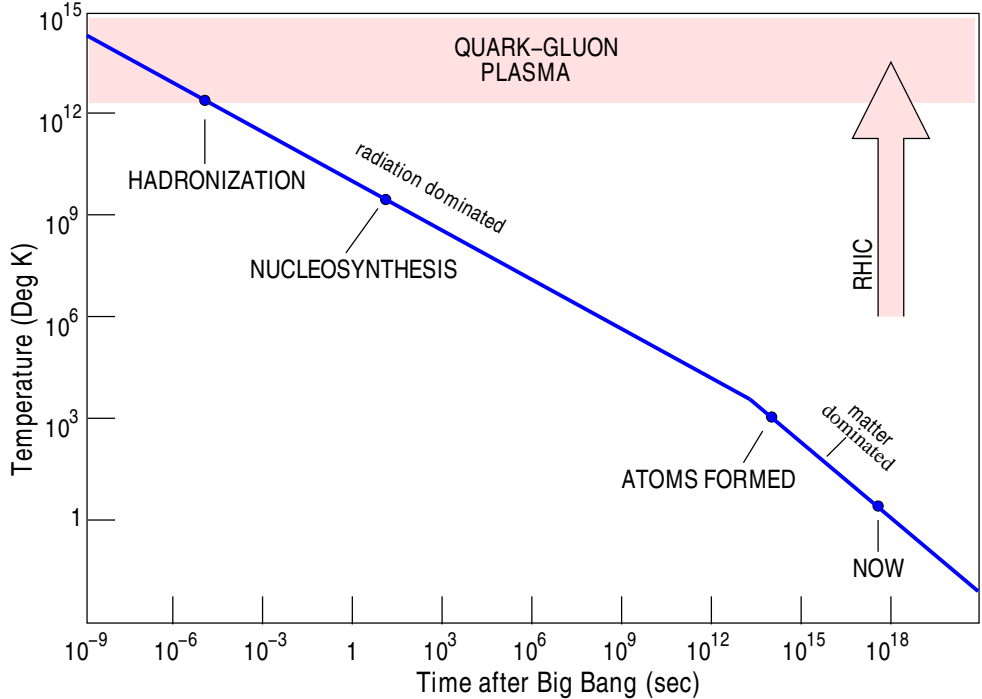}
\caption{Temperature history of the universe.
The Big Bang and the Little Bang.}
\label{T_history}
\end{figure}

Students of high-energy physics would know that the science of the
`small' --- the elementary particle physics --- is deeply intertwined
with the science of the `large' --- cosmology --- the study of the
origin and evolution of the universe.  Figure~\ref{T_history} shows
the temperature history of the universe starting shortly after the Big
Bang. At times $\sim 10 ~\mu s$ after the Big Bang,  
with $T \gsim 200$ MeV,\footnote{In
  comparison, the temperature and time corresponding to the
  electroweak transition were $\sim 200$ GeV and $\sim 10^{-12}$ s,
  respectively.  Note 1 MeV $\simeq 10^{10}$ K.}
the universe was in the state of QGP, and the present-day experiments 
which collide
two relativistic heavy ions --- the Little Bang --- try to recreate
that state of matter in the laboratory for a brief period of time.

\begin{figure}[htbp]
\begin{minipage}{0.4\textwidth}
\center
\includegraphics[scale=0.6]{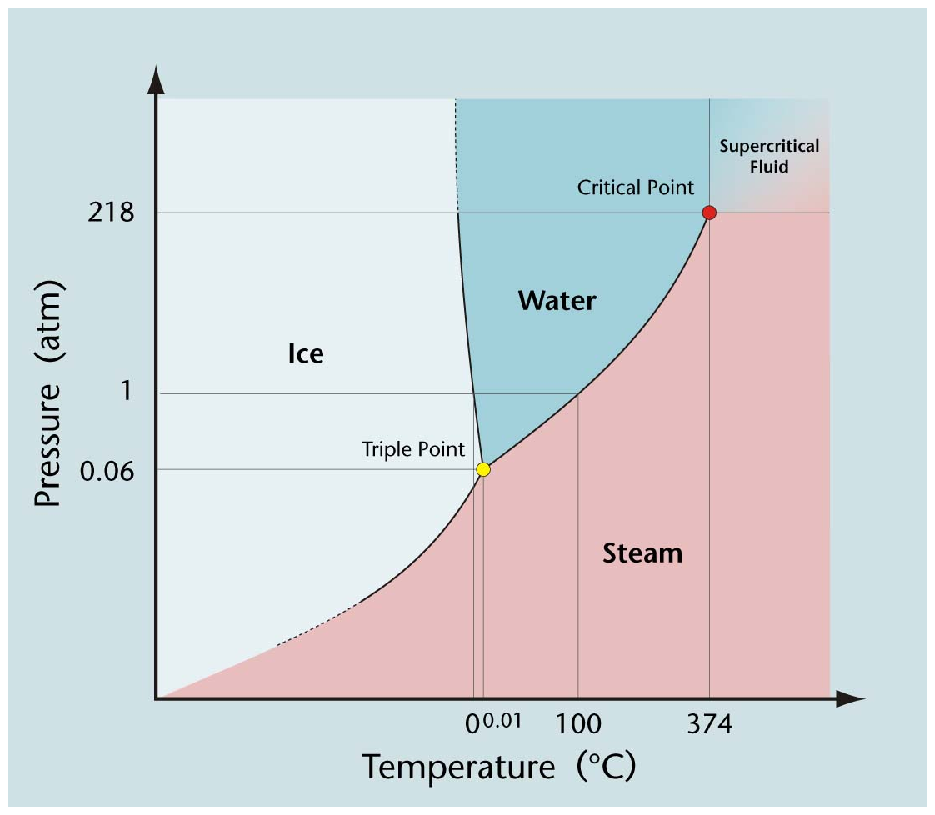}
\end{minipage}
\begin{minipage}{0.6\textwidth}
\center
\includegraphics[scale=0.4]{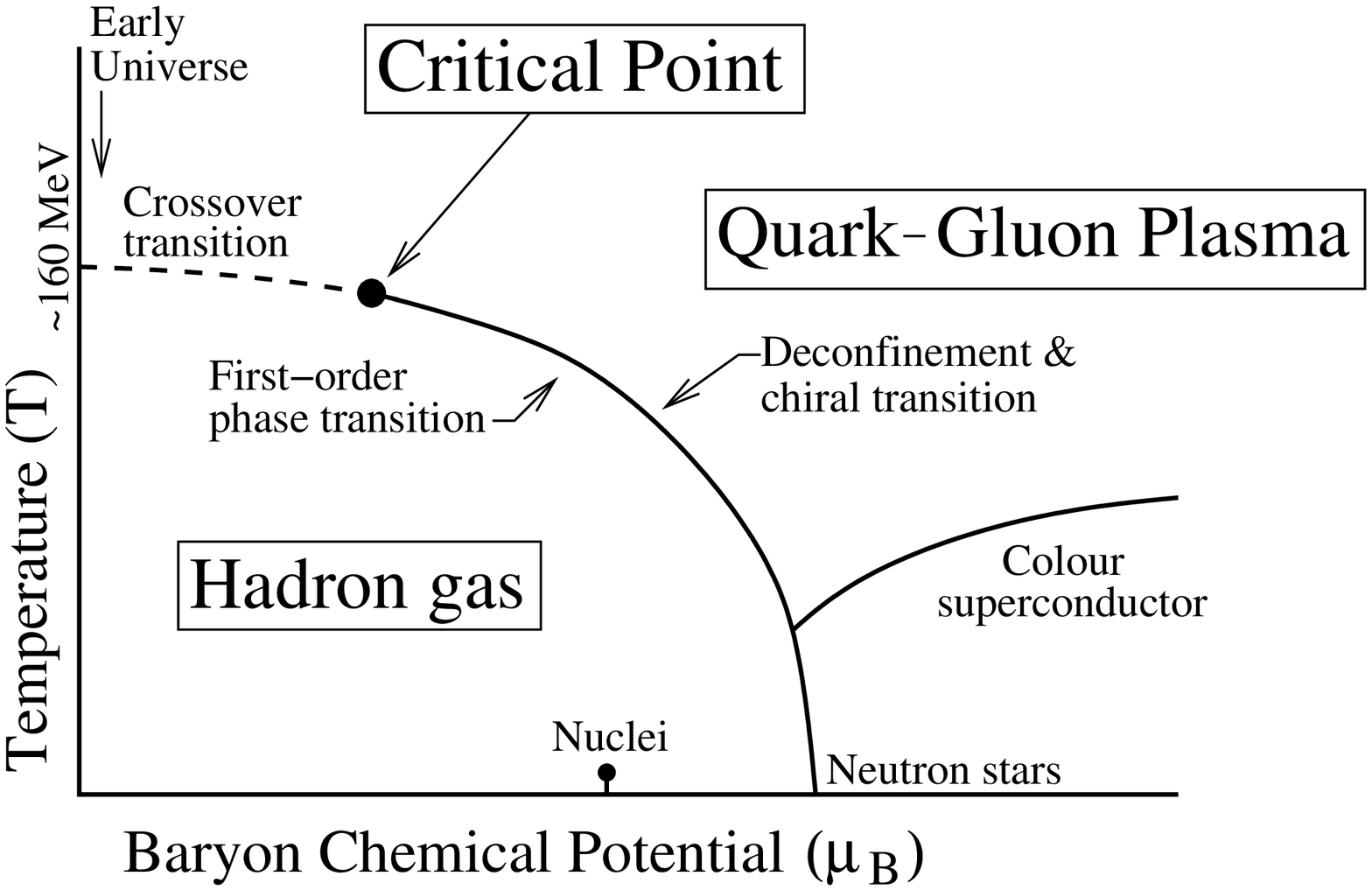}
\end{minipage}
\caption[]{\small (a) Phase diagram of water \cite{langlo} 
and (b) QCD phase diagram
}
\label{phdiags}
\end{figure}

Recall the phase diagram (pressure vs temperature) of water,
Fig.~\ref{phdiags}(a). It shows three broad regions separated by phase
transition lines, the triple point where all three phases coexist, and
the critical point where the vapour pressure curve terminates and two
distinct coexisting phases, namely liquid and gas, become
identical. All these features are well-established experimentally to a
great accuracy. In contrast the QCD phase diagram
(Fig.~\ref{phdiags}(b)) is known only schematically, except for the
lattice QCD predictions at vanishing or small $\mu_B$, in particular 
the prediction of a crossover transition around $T \sim 150$-170
MeV \cite{Borsanyi:2010cj,Borsanyi:2013bia} for vanishing $\mu_B$. 
As arguments based
on a variety of models indicate a first-order phase transition as a
function of temperature at finite $\mu_B$, one expects the phase
transition line to end at a critical point. The existence of the
critical point, however, is not established experimentally. Apart from the
region of hadrons at the low enough $T$ and $\mu_B$, and the region of
quarks and gluons at high $T$ and $\mu_B$, there is also
a region characterized by colour superconductivity, at high $\mu_B$
and low $T$ \cite{Rajagopal:2000wf,Alford:2007xm,Anglani:2013gfu}.
However, precise boundaries separating these regions are not known
experimentally. Actually the QCD phase diagram may be richer than what
is shown in Fig. \ref{phdiags}(b) \cite{McLerran:2007qj}. Before we 
proceed further, a precise
definition of QGP is in order. We follow the definition proposed by
the STAR collaboration at RHIC: Quark-Gluon Plasma is defined as a
(locally) thermally equilibrated state of matter in which quarks and
gluons are deconfined from hadrons, so that they propagate over
\emph{nuclear}, rather than merely \emph{nucleonic}, volumes
\cite{Adams:2005dq}. Note the two essential ingredients of this
definition, (a) the constituents of the matter should be quarks and
gluons, and (b) the matter should have attained
(local)\footnote{Unlike a system in \emph{global} equilibrium, here
  temperature and chemical potential may depend on space-time
  coordinates.} thermal equilibrium. Any claim of discovery of QGP can
follow only after these two requirements are shown to be fulfilled
unambiguously.

The big idea thus is to map out (quantitatively) the QCD phase diagram
\cite{Rajagopal:1999cp}. The main theoretical tool at our disposal is,
of course, the lattice QCD. Although it allows first-principle
calculations, it has technical difficulties for non-vanishing $\mu_B$
or $n_B$. We also have various effective theories and phenomenological
models which indeed are the basis of the schematic phase diagram of
QCD shown in Fig. \ref{phdiags}(b).
Experimental tools available to us are the relativistic heavy-ion
colliders such as those at BNL and CERN, and the upcoming lower-energy
facilities namely Facility for Antiproton and Ion Research (FAIR) at
GSI and Nuclotron-based Ion Collider fAcility (NICA) at JINR.
Apart from these terrestrial facilities, astronomy of neutron stars
can also throw light on the low $T$ and high $n_B$ region of the QCD
phase diagram.

\begin{figure}[htbp]
\begin{minipage}{0.5\textwidth}
\center
\includegraphics[scale=0.41]{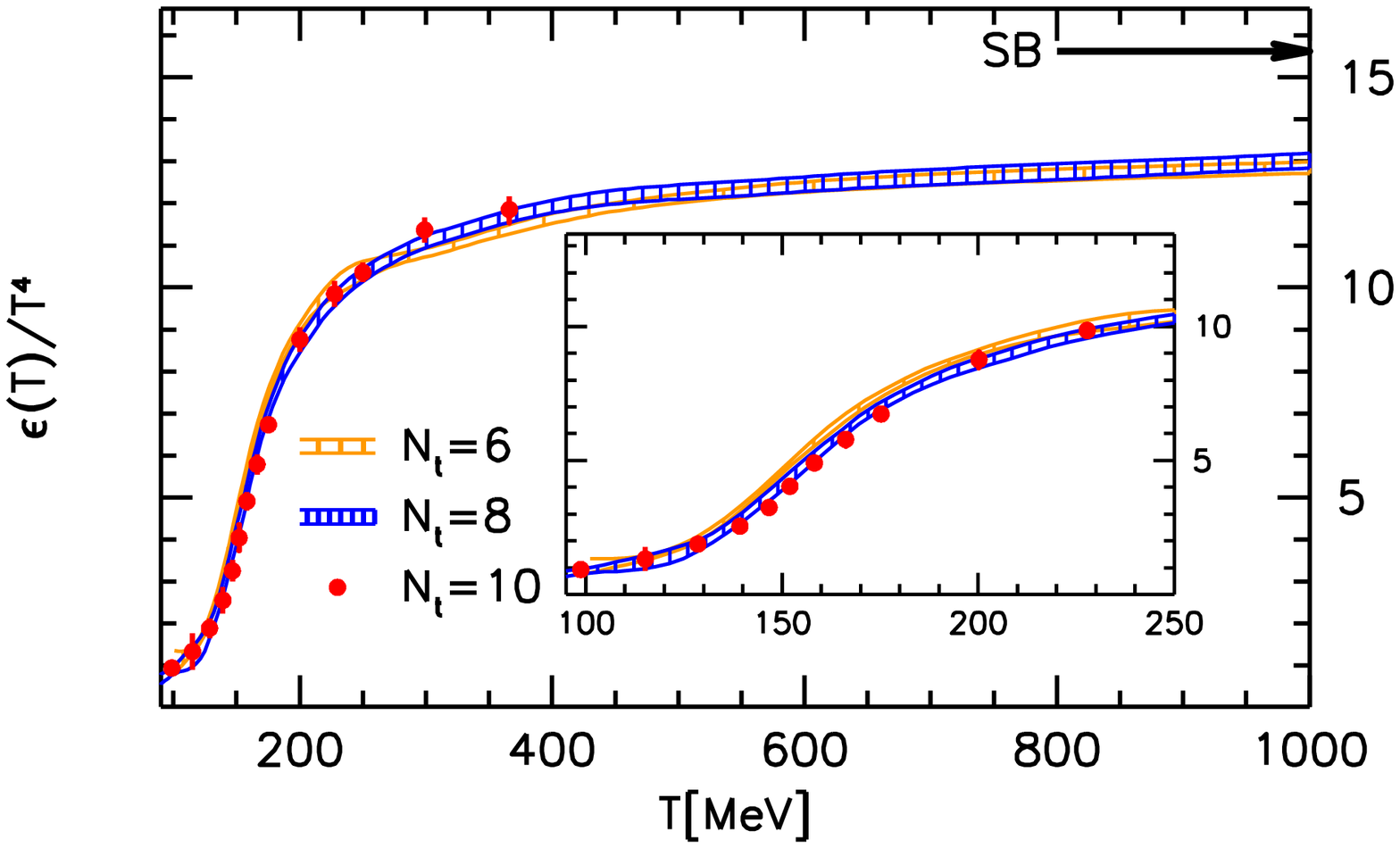}
\end{minipage}
\begin{minipage}{0.5\textwidth}
\center
\includegraphics[scale=0.41]{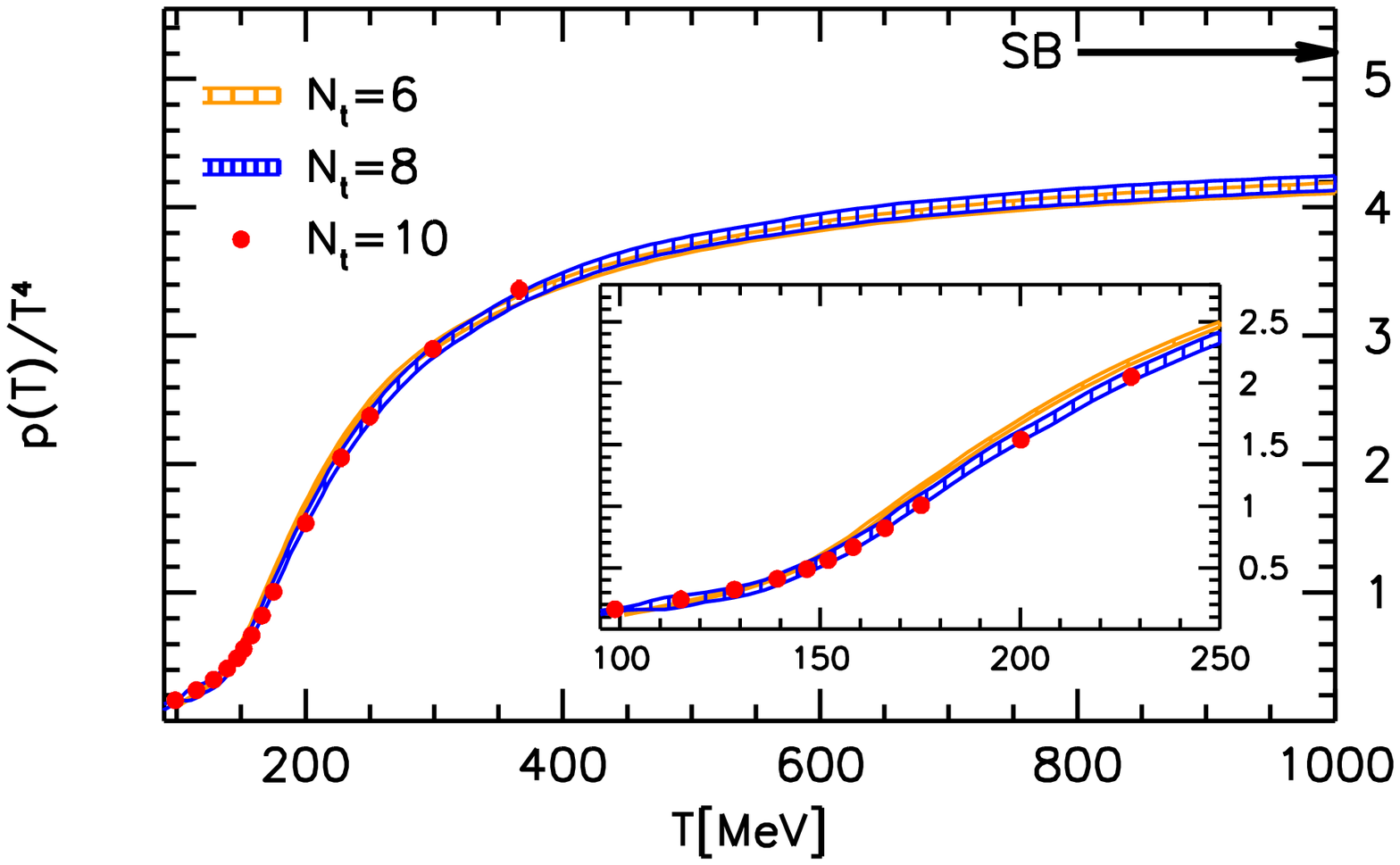}
\end{minipage}\vspace{-3cm}
\caption{Energy density and pressure normalized by $T^4$ as a function
  of temperature ($T$) on $N_t$ = 6, 8 and 10 lattices. $N_t$ is the
  number of lattice points in the temporal direction. The
  Stefan-Boltzmann (SB) limits are indicated by arrows. Figure from
  \cite{Borsanyi:2010cj}; see also \cite{Borsanyi:2013bia}.}
\label{lqcd}
\end{figure}

Figure \ref{lqcd} shows the lattice results for the QCD equation of
state (EoS) at vanishing chemical potential in the temperature range
$100~ {\textmd MeV} \lsim T \lsim 1000$ MeV for physical light and
strange quark masses $m_{u,d,s}$. 
Note that both energy density ($\epsilon$) and
pressure ($P$) rise rapidly around $T=160$ MeV, indicating an increase in
entropy or the number of degrees of freedom. This is consistent with
the deconfinement transition with a concomitant release of the
partonic degrees of freedom. The rise of $P$ is less rapid than that
of $\epsilon$ as expected: the square of the speed of sound
$c_s^2=\partial P/\partial \epsilon$ cannot exceed unity. Note also
that in the limit of high $T$, the EoS approaches the form
$\epsilon=3P$ expected of massless particles. However, $\epsilon$ is
significantly less than $\epsilon_{SB}$ showing that the system is far
from being in an ideal gaseous state. Lattice results indicate that
the transition at vanishing $\mu_B$ is merely an analytic crossover.
Although there is no strict phase transition, it is common to use the
words confined and deconfined phases to describe the low- and
high-temperature regimes. For a recent review of the lattice QCD at
non-zero temperature, see \cite{Petreczky:2012rq}.

An ultrarelativistic heavy-ion collision (URHIC) of two (identical)
Lorentz-contracted\footnote{No matter how high the incoming kinetic
  energy and hence the Lorentz contraction factor is, the limiting
  thickness of the nucleus is $\sim 1$ fm due to the so-called wee
  partons \cite{Bjorken:1976mk}.} nuclei is thought to proceed as
follows. Each incoming nucleus can be looked upon as a coherent
\cite{yagi} cloud of partons (more precisely, a
colour-glass-condensate (CGC) plate \cite{Gelis:2010nm}). The
collision results in shattering of the two CGC plates. A significant
fraction of the incoming kinetic energy is deposited in the central
region leading to a high-energy-density fireball (more precisely, a
highly non-equilibrium state called glasma \cite{Gelis:2010nm}). This
is still a coherent state and liberation of partons from the glasma
takes a finite amount of (proper) time (a fraction of a
fm$/c$). Subsequently collisions among partons lead to a nearly
thermalized (local thermalization!) state called QGP. This happens at
a time of the order of 1 fm$/c$ --- a less understood aspect of the
entire process. Due to near thermalization, the subsequent evolution
of the system proceeds as per relativistic imperfect fluid
dynamics. This involves expansion, cooling, and dilution. Eventually
the system hadronizes. Hadrons continue to collide among themselves
elastically which changes their energy-momenta, as well as
inelastically which alters abundances of individual species. Chemical
freezeout occurs when inelastic processes stop. Kinetic freezeout
occurs when elastic scatterings too stop. These late stages of
evolution when the system is no longer in local equilibrium are
simulated using the relativistic
kinetic theory framework. Hadrons decouple from
the system approximately 10-15 fm$/c$ after the collision and travel
towards the surrounding detectors. From the volume of experimental
data thus collected one has to establish whether QGP was formed and if
so, extract its properties.

After years of work a Standard Model of URHICs has emerged: The
initial state is constructed using either the Glauber model
\cite{Miller:2007ri} or one of the models implementing ideas
originating from CGC \cite{Kharzeev:2000ph}; 
for a recent review see \cite{Albacete:2014fwa}. The intermediate
evolution is considered using some version of the
M\"uller-Israel-Stewart-like theory \cite{Muller:1967zza,Israel:1979wp} of
causal relativistic imperfect fluid dynamics, together with a
QCD equation of state spanning partonic and hadronic phases
\cite{Huovinen:2009yb}.
The end evolution
of the hadron-rich medium
leading to a freezeout uses the Boltzmann equation in the relativistic
transport theory \cite{Song:2011hk}. 
The final state consists of thousands of particles
(mesons, baryons, leptons, photons, light nuclei). Detailed measurements
(single-particle inclusive, two- and multi-particle correlations,
etc.) are available, spanning the energy range from SPS to RHIC to LHC, 
for various colliding nuclei, centralities, (pseudo)rapidities, and
transverse momenta. The aim is to achieve a quantitative
understanding of the thermodynamic and transport
properties of QGP, \eg its EoS, its transport coefficients
(shear and bulk viscosities, diffusivity, conductivity),
etc. The major hurdles in this endeavour are an inadequate knowledge
of the initial state and event-to-event fluctuations at nucleonic 
and sub-nucleonic levels in the initial state.

% ---------------------------------------------------------------------------
\section{Two most important observables}

Elliptic flow and jet quenching are arguably the two most important
observables in this field. Observation of an elliptic flow almost as
large as that predicted by ideal (\ie equilibrium) hydrodynamics led
to the claim of formation of an almost perfect fluid at RHIC
\cite{Gyulassy:2004zy}. A natural explanation of the observed jet
quenching is in terms of a dense and coloured (hence partonic, not
hadronic) medium that is rather opaque to high-momentum
hadrons. Recall the definition of QGP given in section 1. The two
essential requirements mentioned there seem to be fulfilled
considering these two observations together.

\begin{figure}[htbp]
\centering\includegraphics[width=.5\linewidth]{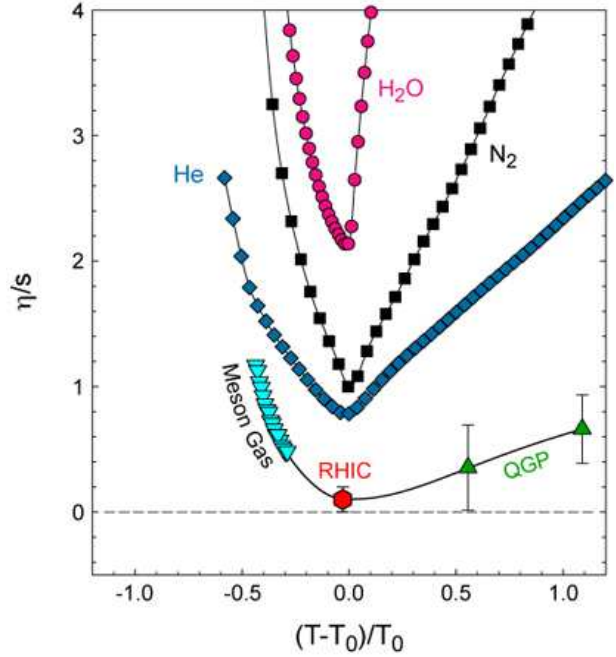}
\caption{Constant pressure ($P_{critical}$)
curves for (shear viscosity/entropy
  density) vs temperature. $T_0$ is the critical temperature of the
  liquid-gas phase transition. Points labelled Meson Gas are based on
  chiral perturbation theory and have $50 \%$ errors (not
  shown). Points labelled QGP are based on lattice QCD
  simulations. Figure from \cite{Lacey:2006bc}.}
\label{lacey}
\end{figure}

Before I discuss these two observations in detail, let me explain what
is meant by an almost perfect fluid. Air and water are the two most
common fluids we encounter. Which of them is more viscous? Water has a
higher coefficient of shear viscosity ($\eta$) than air, and appears
more viscous. But that is misleading. To compare different fluids, one
should consider their kinematic viscosities defined as $\eta/\rho$
where $\rho$ is the density. Air has a higher kinematic viscosity and
hence is actually more viscous than water! Relativistic analogue of
$\eta/\rho$ is the dimensionless ratio $\eta/s$ where $s$ is the
entropy density. Scaling by $s$ is appropriate because number density
is ill-defined in the relativistic case. Figure \ref{lacey} shows
constant-pressure ($P_{critical}$) curves for $\eta/s$ as a function
of temperature for various fluids, namely water, nitrogen, helium, and
the fluid formed at RHIC. All fluids show a minimum at the critical
temperature, and among them the RHIC fluid has the lowest $\eta/s$,
even lower than that of helium. Hence it is the most perfect fluid
observed so far\footnote{More recently, trapped ultracold atomic
  systems are also shown to have $\eta/s$ much smaller than that for
  helium \cite{Schafer:2009dj}.}. For water, nitrogen, and helium,
points to the left (right) of the minimum refer to the liquid
(gaseous) phase.  As $T$ rises, $\eta/s$ for these liquids drops, attains a
minimum at the critical temperature $T_0$, and then in the gaseous phase 
it rises. This is
because liquids and gases transport momentum differently
\cite{Csernai:2006zz}. RHIC fluid is an example of a strongly coupled
quantum fluid and has been called sQGP to distinguish it from weakly
coupled QGP or wQGP expected at extremely high temperatures.
Interestingly, the liquid formed at RHIC and LHC cools into a 
(hadron resonance) gas!

% ---------------------------------------------------------------------------

\subsection{Elliptic flow}
\begin{figure}[htbp]
\begin{minipage}{0.5\textwidth}
\center
\includegraphics[scale=.5]{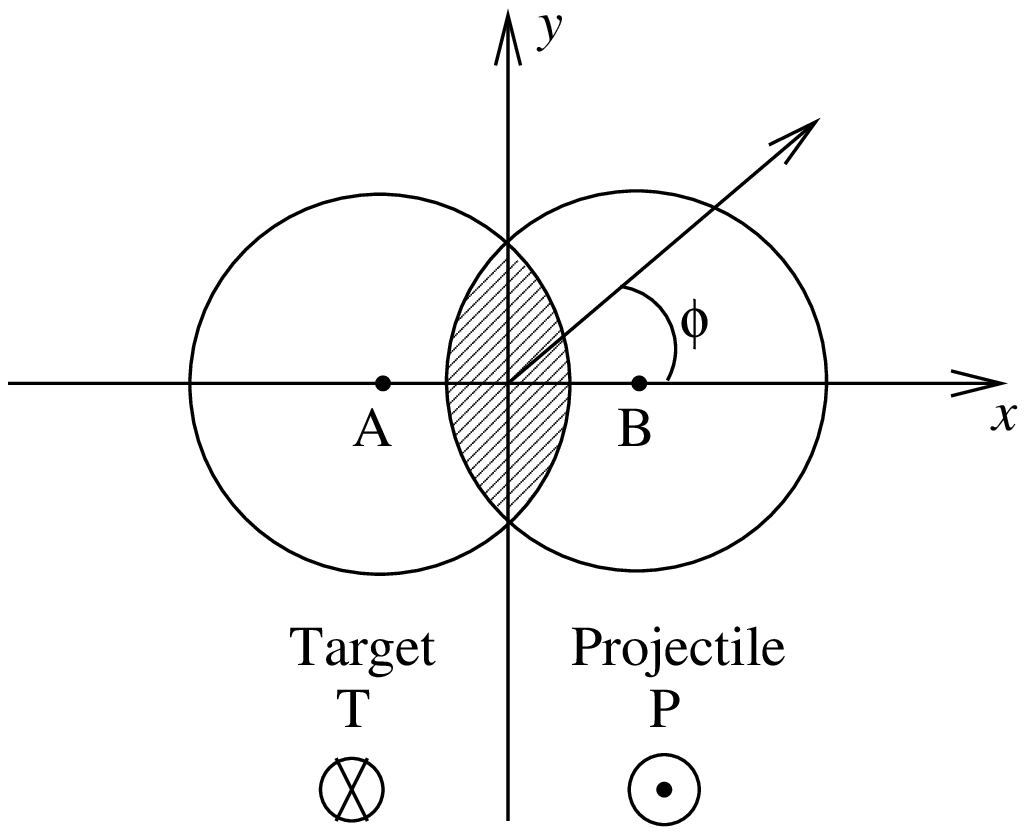}
\end{minipage}
\begin{minipage}{0.5\textwidth}
\center
\includegraphics[scale=.6]{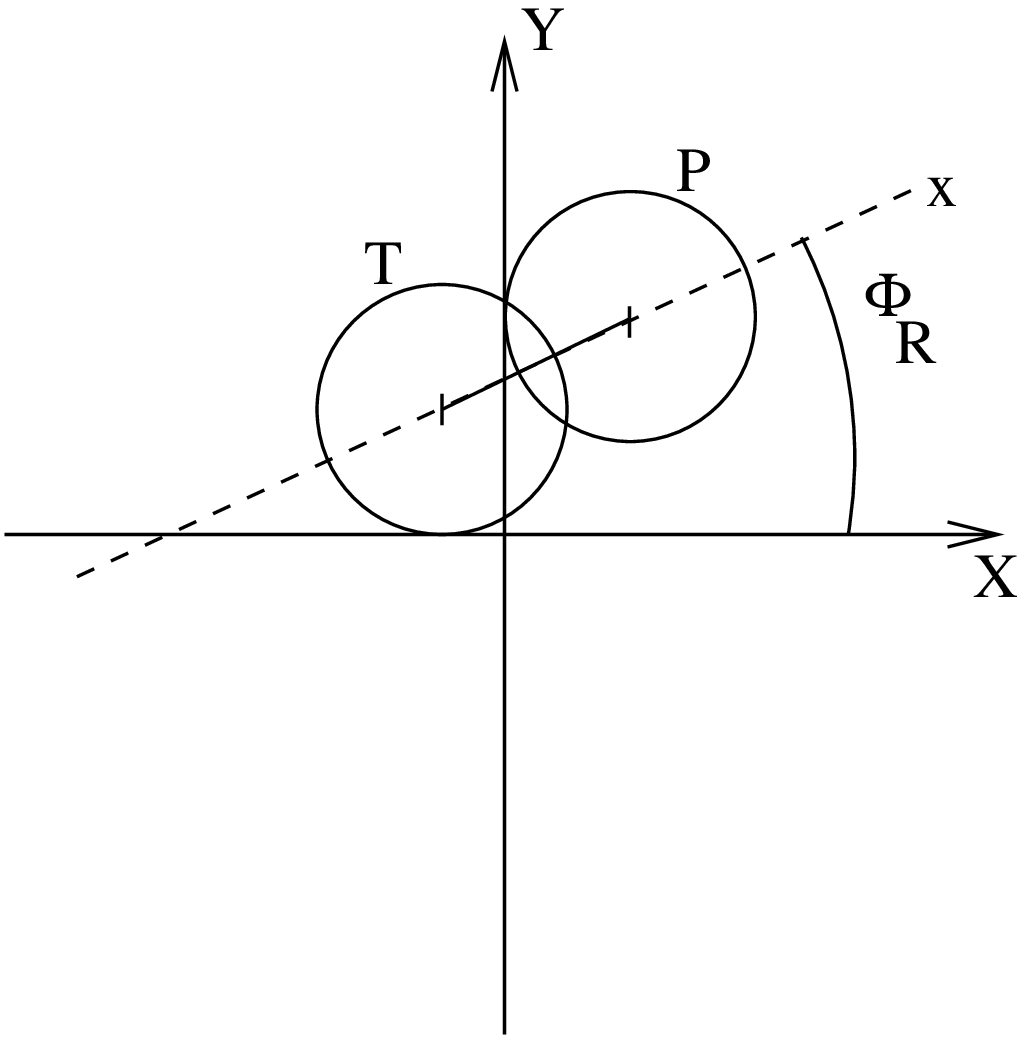}
\end{minipage}
\caption{(a) Non-central collision of two nuclei. Collision or beam
  axis is perpendicular to the plane of the figure. Impact parameter
  $b$ = length AB. $z$ is the longitudinal direction, $xy$ is the
  transverse or azimuthal plane, $xz$ is the reaction plane, and
  $\phi$ is the azimuthal angle of one of the outgoing particles. The
  shaded area indicates the overlap zone. For a central or head-on
  collision ($b=0$) the reaction plane cannot be defined. (b) $XYZ$
  are the lab-fixed axes. $\Phi_R$ is the reaction-plane angle.}
\label{colli}
\end{figure}

Consider a non-central (\ie non-zero impact parameter) collision of
two identical spherical nuclei travelling in opposite directions; see
Fig. \ref{colli}(a). In an actual experiment the magnitude and
orientation of the impact parameter vector fluctuate from event to
event (Fig. \ref{colli}(b)) and are unknown. This initial geometry can
potentially affect the distribution of particles in the final state
--- in particular, in the transverse plane. In order to capture this
physics in terms of a few parameters, the triple differential
invariant distribution of particles emitted in the final state is
Fourier-decomposed as follows \cite{Voloshin:1994mz}
\begin{equation}
E\frac{d^3N}{d^3p} = \frac{d^3N}{p_Tdp_Tdyd\phi}
= \frac{d^2N}{p_Tdp_Tdy}\frac{1}{2\pi} \left[ 1+ \sum_{n=1}^\infty
2 v_n \cos n(\phi-\Phi_R) \right] ,
\label{triple}
\end{equation}
where $p_T$ is the transverse momentum, $y$ the rapidity, $\phi$ the
azimuthal angle of the outgoing particle momentum, and $\Phi_R$ the
reaction-plane angle. 
Sine terms, $\sin n(\phi-\Phi_R)$, are not included in the Fourier
expansion in 
Eq. (\ref{triple}) because they vanish
due to the reflection symmetry with respect to the reaction plane; see
Fig. \ref{colli}. 
The reaction-plane angle $\Phi_R$ which characterizes 
the initial geometry
(Fig. \ref{colli}(b))
is not known, and is
estimated using the transverse distribution of particles in the final
state. The estimated reaction plane is called the event plane. 
The
leading term in the square brackets in Eq. (\ref{triple}) represents
the azimuthally symmetric radial flow. The first two harmonic
coefficients $v_1$ and $v_2$ are called directed and elliptic
flows, respectively\footnote{To understand this
  nomenclature, make polar plots of $r=(1+2v_n \cos n\phi)$ for a
  small positive value of $v_n$.}. We have
\begin{equation}
v_n(p_T,y) = \mean{\cos [n(\phi-\Phi_R)]}
= \frac{\int^{2\pi}_0 d\phi \cos[n(\phi-\Phi_R)] \frac{d^3N}
{p_T dp_T dy d\phi}} {\int^{2\pi}_0 d\phi
\frac{d^3N}{p_T dp_T dy d\phi}} .
\end{equation}
The average is taken in the $(p_T,y)$ bin under consideration. After
taking the average over all particles in an event, average is then
taken over all events in a centrality class\footnote{Centrality of a
  $AA$ collision is determined making use of its tight correlation
  with the charged-particle multiplicity or transverse energy at
  mid-rapidity, which in turn are anti-correlated with the energy
  deposited in the Zero Degree Calorimeters.}. For a central collision
the azimuthal distribution is isotropic, and hence $v_n=0$, \ie only
the radial flow survives.
For a review of the methods used for analyzing anisotropic flow in
relativistic heavy-ion collisions, and interpretations and
uncertainties in the measurements, see
\cite{Poskanzer:1998yz,Voloshin:2008dg}.

In a non-central collision, the initial state is characterized by a
\emph{spatial anisotropy} in the azimuthal plane
(Fig.~\ref{colli}). Consider particles in the almond-shaped overlap
zone. Their initial momenta are predominantly longitudinal. Transverse
momenta, if any, are distributed isotropically. If these particles do
not interact with each other, the final (azimuthal) distribution too
will be isotropic. On the other hand, if they do interact with each
other frequently and with adequate strength (or cross section), then
the (local) thermal equilibrium is likely to be reached. Once that
happens, the system can be described in terms of thermodynamic
quantities such as temperature, pressure, etc. The spatial anisotropy
of the overlap zone ensures anisotropic pressure gradients in the
transverse plane. This leads to a final state characterized by
\emph{momentum anisotropy}, an anisotropic azimuthal distribution of
particles, and hence a nonvanishing $v_n$. Thus $v_n$ is a measure of
the degree of thermalization of the quark-gluon matter produced in a
noncentral heavy-ion collision --- a central issue in this field.
                                   
The anisotropic flow $v_n$ is sensitive to the \emph{early} ($\sim$
fm/$c$) history of the collision: Higher pressure gradients along the
minor axis of the spatially anisotropic source (Fig.~\ref{colli})
imply that the expansion of the source would gradually diminish its
anisotropy, making the flow self-quenching. Thus $v_n$ builds up early 
(\ie when
the anisotropy is significant) and tends to saturate as the anisotropy
continues to decrease.  (This is unlike the radial flow which
continues to grow until freezeout and is sensitive to early- as well
as late-time history of the collision). Thus $v_n$ is a signature of
pressure at early times.

The flow $v_n$ depends on the initial conditions, \ie the beam energy,
the mass number of colliding nuclei, and the centrality of the
collision. It also depends on the species of the particles under
consideration apart from their transverse momentum $(p_T)$ and
rapidity $(y)$ or pseudorapidity $(\eta)$. Using the symmetry of the
initial geometry, one can show that $v_n(y)$ is an even (odd) function
of $y$ if $n$ is even (odd). Hence $v_1(y)$
\begin{figure}[htbp]
\centering\includegraphics[scale=0.7]{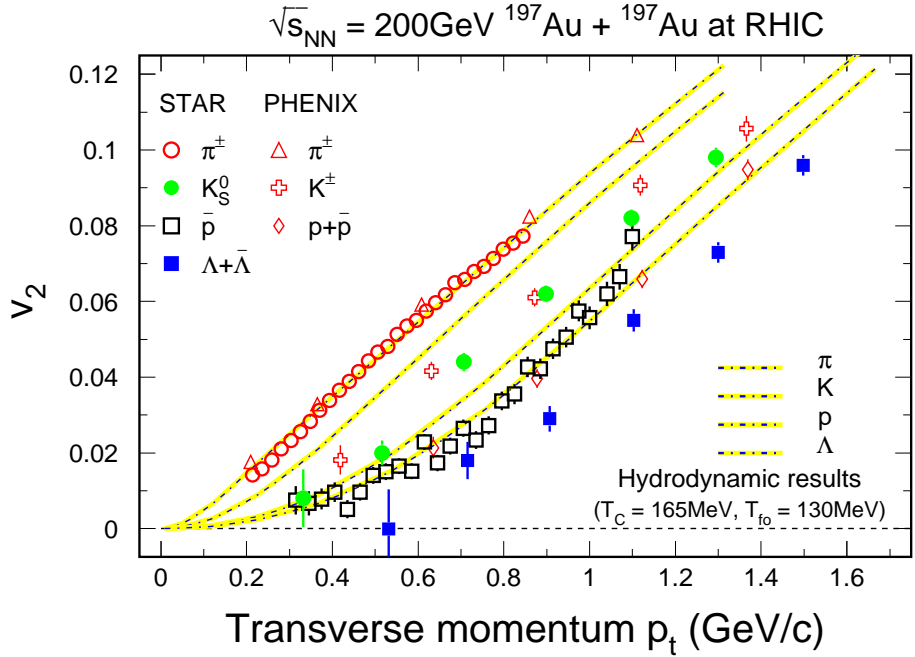}
\caption{Success of ideal hydrodynamics: Minimum-bias elliptic flow
  data for different particle species in comparison with ideal
  hydrodynamics calculations. Figure from \cite{Oldenburg:2004qa}.}
\label{v2}
\end{figure}
vanishes at mid-rapidity. At RHIC energies at mid-rapidity, it is the
elliptic flow $v_2$ that plays an important role. Figure~\ref{v2}
shows the $v_2(p_T)$ data at the highest RHIC energy for various
particle species, in broad agreement with the ideal hydrodynamic
calculations. As stated before, this success of the ideal
hydrodynamics led to the claim of formation of an almost perfect fluid
at RHIC.

\emph{Extraction of} $\eta/s$: Introduction of shear viscosity tends
to reduce the elliptic flow, $v_2$, with respect to that for an ideal
fluid: a particle moving in the reaction plane (Fig.~\ref{colli}(a))
being faster experiences a greater frictional force compared with a
particle moving out of the plane thereby reducing the azimuthal
anisotropy and hence $v_2$. This fact has been used to place an upper
limit on the value of $\eta/s$ of the RHIC fluid. A more precise
determination is hindered by ambiguities in the knowledge of the
initial state. Event-to-event fluctuations give rise to `new' flows
and observables which help constrain the $\eta/s$ further.

% ----------------------------------------------------------------------------

\subsubsection{Event-to-event fluctuations}

The discussion above was somewhat idealistic because we assumed smooth
initial geometry: Energy (or entropy) density $\epsilon(x,y)$ 
$(\textrm{or}~s(x,y))$ in the shaded area in Fig. \ref{colli}(a) was a smooth
function of $x,y$ because it was assumed to result from the overlap of
two smooth Woods-Saxon nuclear density distributions. However, the
reality is not so simple, \ie the initial geometry is not smooth.

\begin{figure}[htbp]
\centering\includegraphics[width=0.5\linewidth]{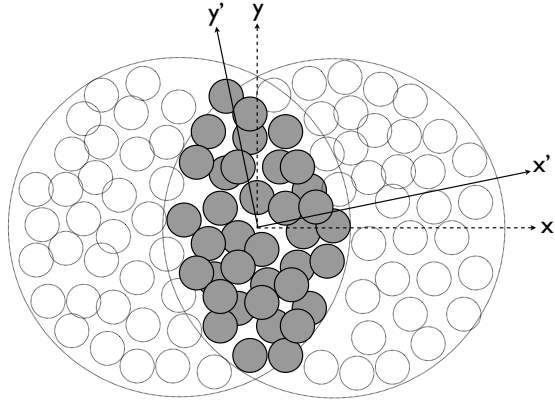}
\caption{`Snapshot' of nucleon positions at the instant of
  collision. Due to event-to-event fluctuations, the overlap zone
  could be shifted and tilted with respect to the $(x,y)$
  frame. $x'y'$: principal axes of inertia. Figure from
  \cite{Alver:2008zza}.}
\label{eefluct}
\end{figure}

\begin{figure}[htbp]
\begin{minipage}{0.5\textwidth}
\center
\includegraphics[scale=.5]{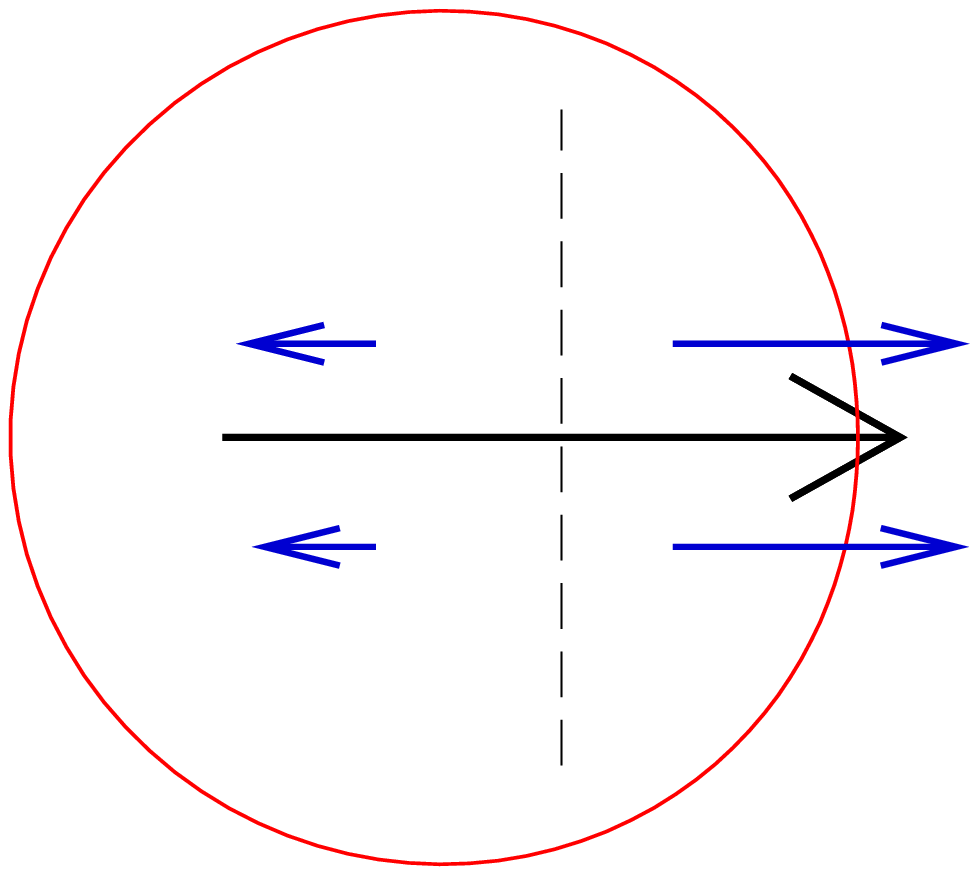}
\end{minipage}
\begin{minipage}{0.5\textwidth}
\center
\includegraphics[scale=.35]{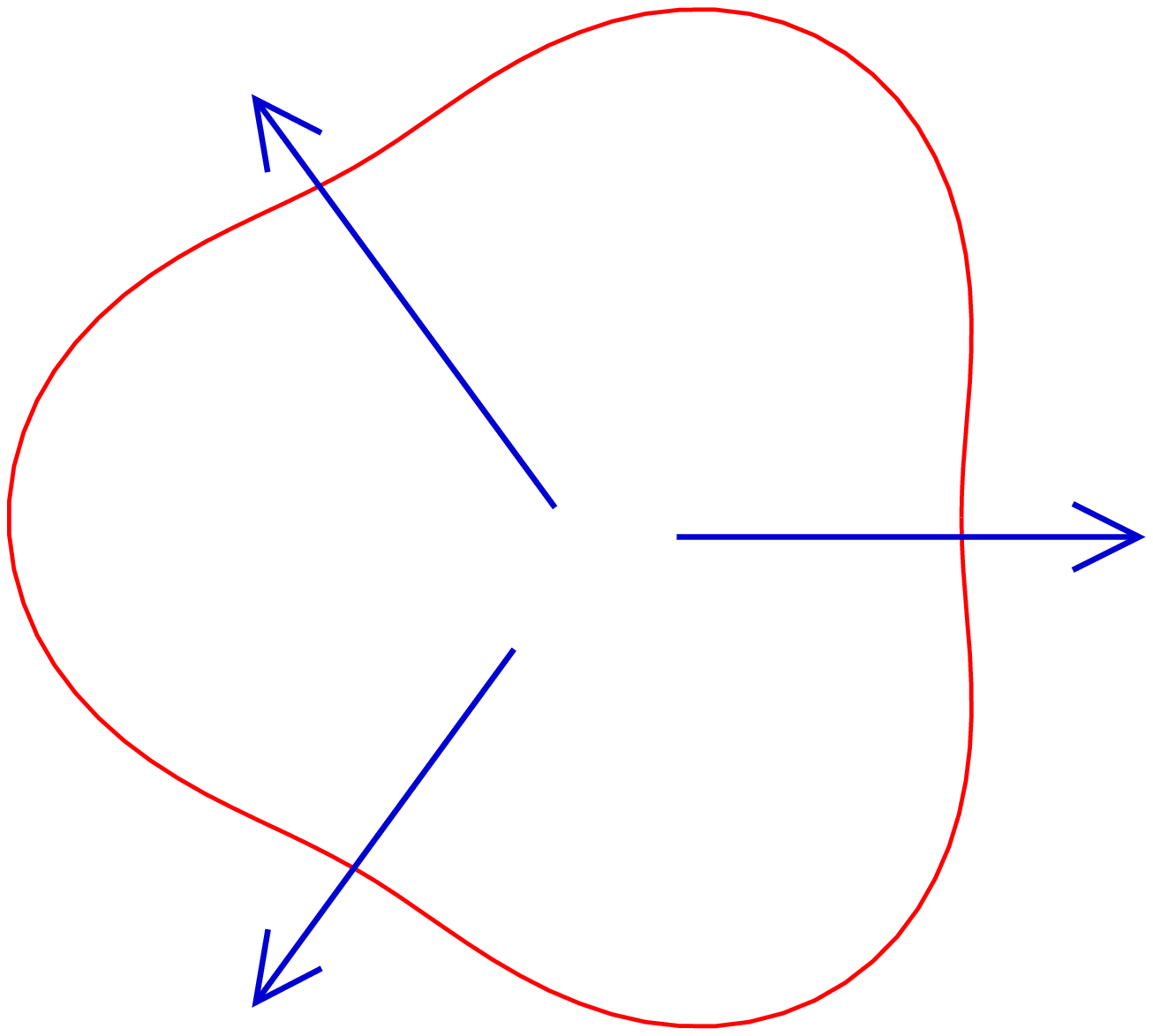}
\end{minipage}
\caption {(a) Dipole asymmetry giving rise to a dipolar flow
  $v_1(p_T,y)$. The cross indicates the centre of entropy (analogous
  to the centre of mass) and the large arrow indicates the orientation
  of the dipole. (b)Triangularity giving rise to a triangular flow
  $v_3(p_T,y)$. Figure from \cite{Teaney:2010vd}.}
\label{newflow}
\end{figure}

In relativistic heavy-ion collisions, the collision time-scale is so
short that each incoming nucleus sees nucleons in the other nucleus in
a frozen configuration. Event-to-event fluctuations in nucleon ($N$)
positions (and hence in $NN$ collision points) result in an overlap
zone with inhomogeneous energy density and a shape that fluctuates
from event to event, Fig. \ref{eefluct}. This necessitates that
the ``sine terms'' are also included in the Fourier expansion in
Eq. (\ref{triple}). Equivalently, one writes
\begin{equation}
E\frac{d^3N}{d^3p} = \frac{d^3N}{p_Tdp_Tdyd\phi}
= \frac{d^2N}{p_Tdp_Tdy}\frac{1}{2\pi} \left[ 1+ \sum_{n=1}^\infty
2 v_n \cos n(\phi-\Psi_n) \right].
\label{triple1}
\end{equation}
Thus each harmonic $n$ may have its own reference angle $\Psi_n$ in the
transverse plane.
Traditional hydrodynamic
calculations do not take these event-to-event fluctuations into
account. Instead of averaging over the fluctuating initial conditions and then
evolving the resultant smooth distribution, one needs to perform
event-to-event hydrodynamics calculations first and then average over
all outputs. This is done in some of the recent hydrodynamic
calculations. They also incorporate event-to-event fluctuations at
the sub-nucleonic level.
Fluctuating initial geometry results in `new' (rapidity-even) flows
(Fig.~\ref{newflow}). The rapidity-even dipolar flow shown in 
Fig.~\ref{newflow}(a) is not to be confused with the rapidity-odd directed
flow $v_1(p_T,y)$ resulting from the smooth initial geometry in 
Fig.~\ref{colli}.

For recent reviews of the collective flow, its anisotropies, its
event-to-event fluctuations, and the extraction of the specific shear
viscosity $\eta/s$ of QGP, see \cite{Heinz:2013th,Gale:2013da,
Huovinen:2013wma}.

% ----------------------------------------------------------------------------

\subsection{Jet quenching}

\begin{figure}[htbp]
\begin{minipage}{0.3\textwidth}
\center
\includegraphics[scale=.23]{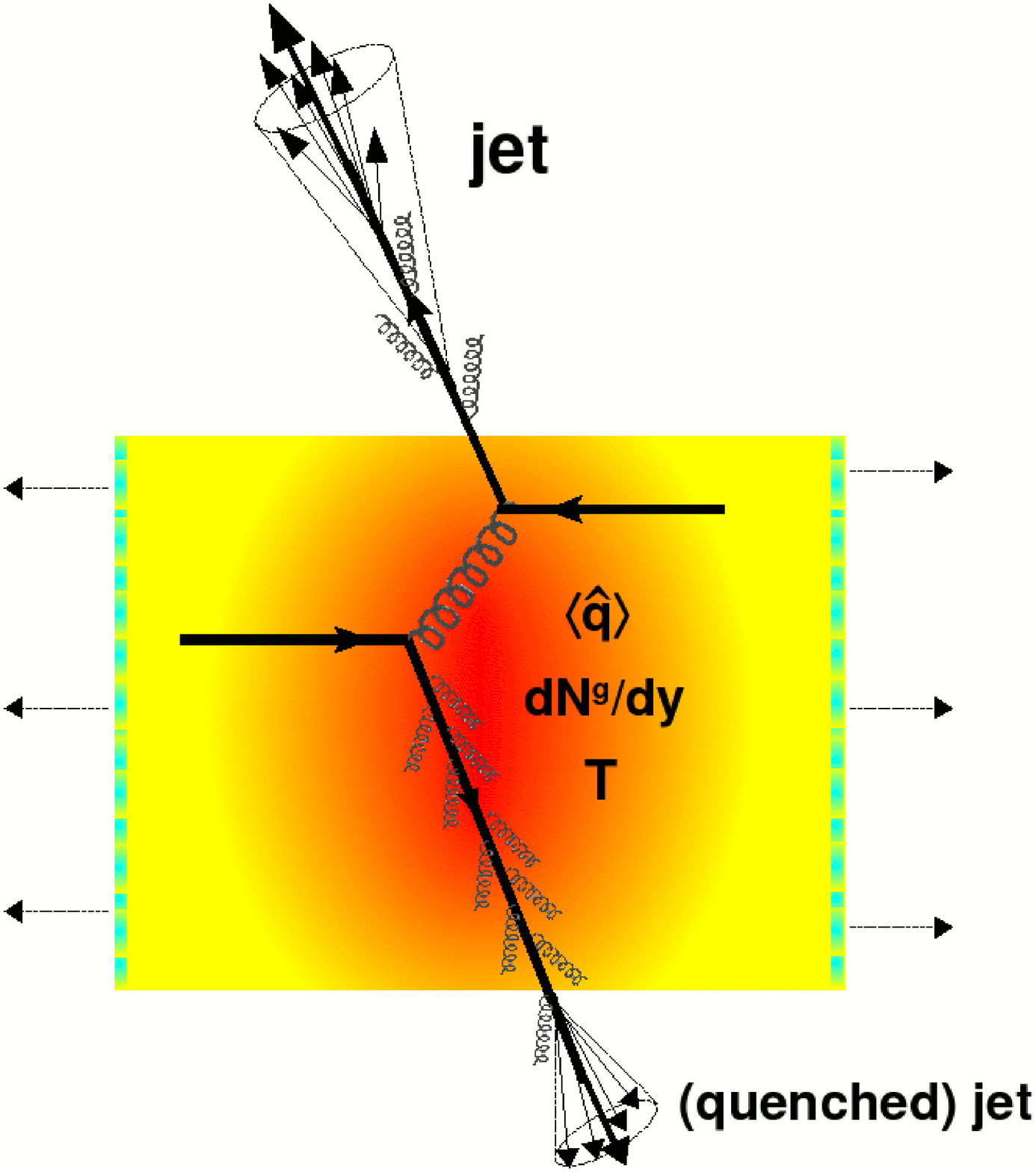}
\end{minipage}
\begin{minipage}{0.7\textwidth}
\center
\includegraphics[scale=0.5]{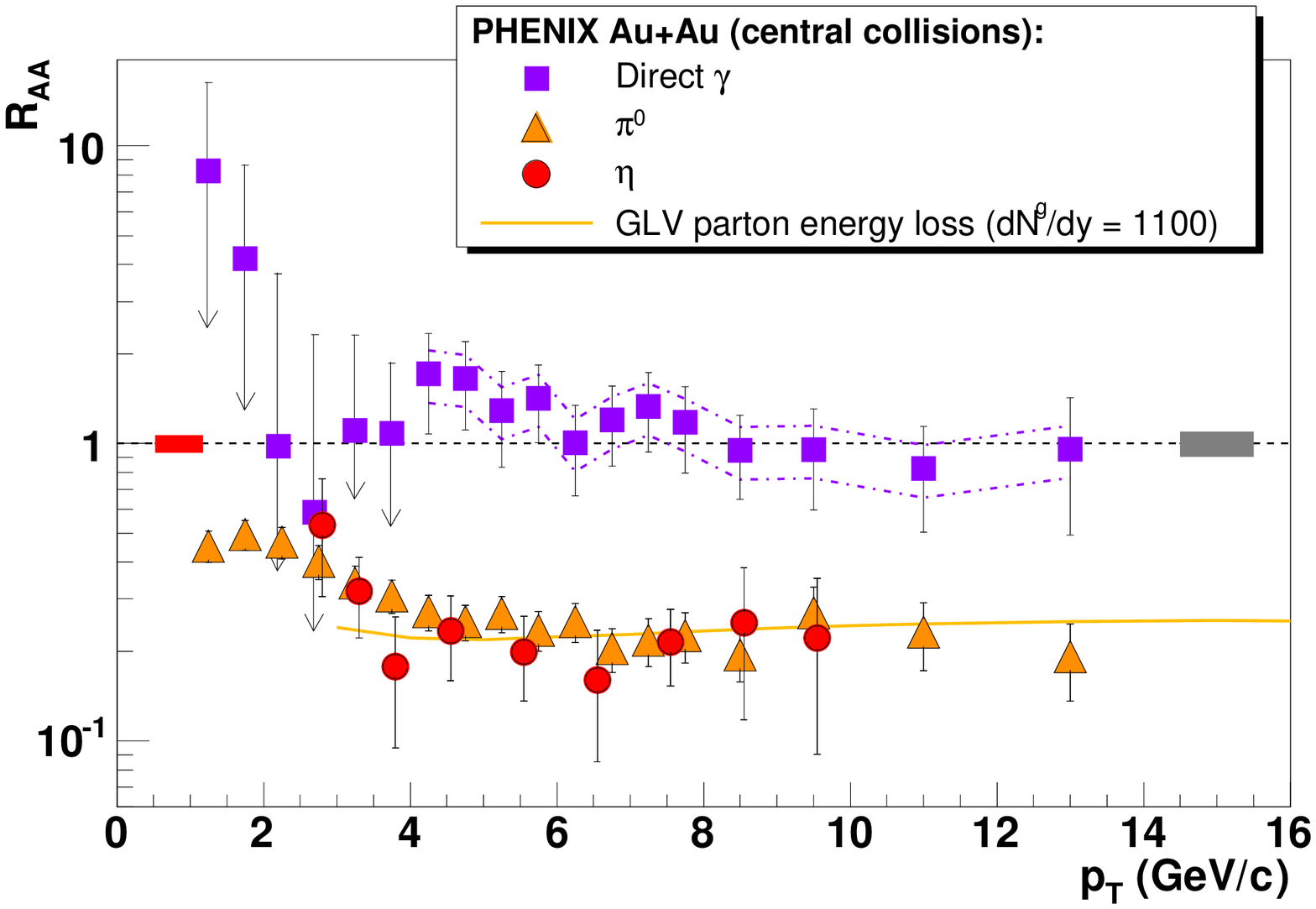}
\end{minipage}
\caption{(a) Back-to-back jets, one produced near the surface of the
  hot and dense medium and the other deep inside. These are called the
  near-side and away-side jets. The latter gets quenched. The medium
  is characterized by its temperature ($T$), gluon number density in
  the rapidity space ($dN^g/dy$), and the transport coefficient or
  jet-quenching parameter ($\hat{q}$). Figure from
  \cite{d'Enterria:2009am}. (b) AuAu central collision data on
  nuclear modification factor $R_{AA}$ as a function of $p_T$, at
  the centre-of-mass energy
  $\sqrt{s_{NN}}=200$ GeV. Dash-dotted lines: theoretical uncertainties in
  the direct photon $R_{AA}$. Solid yellow line: jet-quenching calculation
  of \cite{Vitev:2002pf,Vitev:2004bh} for leading pions in a medium
  with initial effective gluon density $dN^g/dy =1100$. Error bands at
  $R_{AA}=1$ indicate the absolute normalization errors. Figure from
  \cite{Adler:2006bv}.}
\label{jq1}
\end{figure}

Recall the role played by successively higher-energy electron beams,
over many decades in the last century, to unravel the structure of
atoms, nuclei, and protons. Studying the properties of QGP by means of an
external probe is obviously ruled out because of its short ($\sim
10^{-23}$ s) life-time. Instead one uses a hard parton produced
internally during the nucleus-nucleus collision to probe the medium in
which it is produced. Consider, \eg $g+g \rightarrow g+g$ where two
longitudinally moving energetic gluons from the colliding nuclei
interact and produce two gluons at large transverse momenta, which
fragment and emerge as jets of particles. Hard partons are produced
early in the collision: $t \sim 1/Q \sim 1/p_T$, where $Q$ is the
parton virtuality scale, and hence they probe the early stages of the
collision. Moreover, their production rate is calculable in
perturbative QCD. Parton/jet interacts with the medium and loses
energy or gets quenched as it traverses the medium
(Fig.~\ref{jq1}(a)). The amount of energy loss depends among other
things on the path length ($L$) the jet has to travel inside the
medium. Figure~\ref{jq1}(b) shows the data on the nuclear modification
factor, $R_{AA}$, defined schematically as
\begin{equation}
R_{AA}(p_T)=\textrm{Yield ~in~} AA / \mean{N_{coll}} \textrm{Yield~ in~}pp , 
\label{yield}
\end{equation}
where $\mean{N_{coll}}$ is the mean number of nucleon-nucleon
collisions occurring in a single nucleus-nucleus ($AA$) collision,
obtained within the Glauber model \cite{Miller:2007ri}. If the
nucleus-nucleus collision were a simple superposition of
nucleon-nucleon collisions, the ratio $R_{AA}$ would be
unity. Direct-photon production rate is consistent with the
next-to-leading-order (NLO) perturbative QCD (pQCD) calculation and
there is no suppression of the photon yield. However, the yields of
high-$p_T$ pions and etas are suppressed by a factor of $\sim
5$. No such suppression was seen in $d$Au and $p$Pb collisions 
\cite{pPbRAA} (where
QGP is not expected to be formed) thereby ruling out suppression
by cold nuclear matter as the cause.
These observations indicate that the hard-scattered partons lose
energy as they traverse the hot medium and the suppression is thus a
final-state effect.

\begin{figure}[htbp]
\begin{minipage}{0.5\textwidth}
\center
\includegraphics[height=53mm,width=81mm]{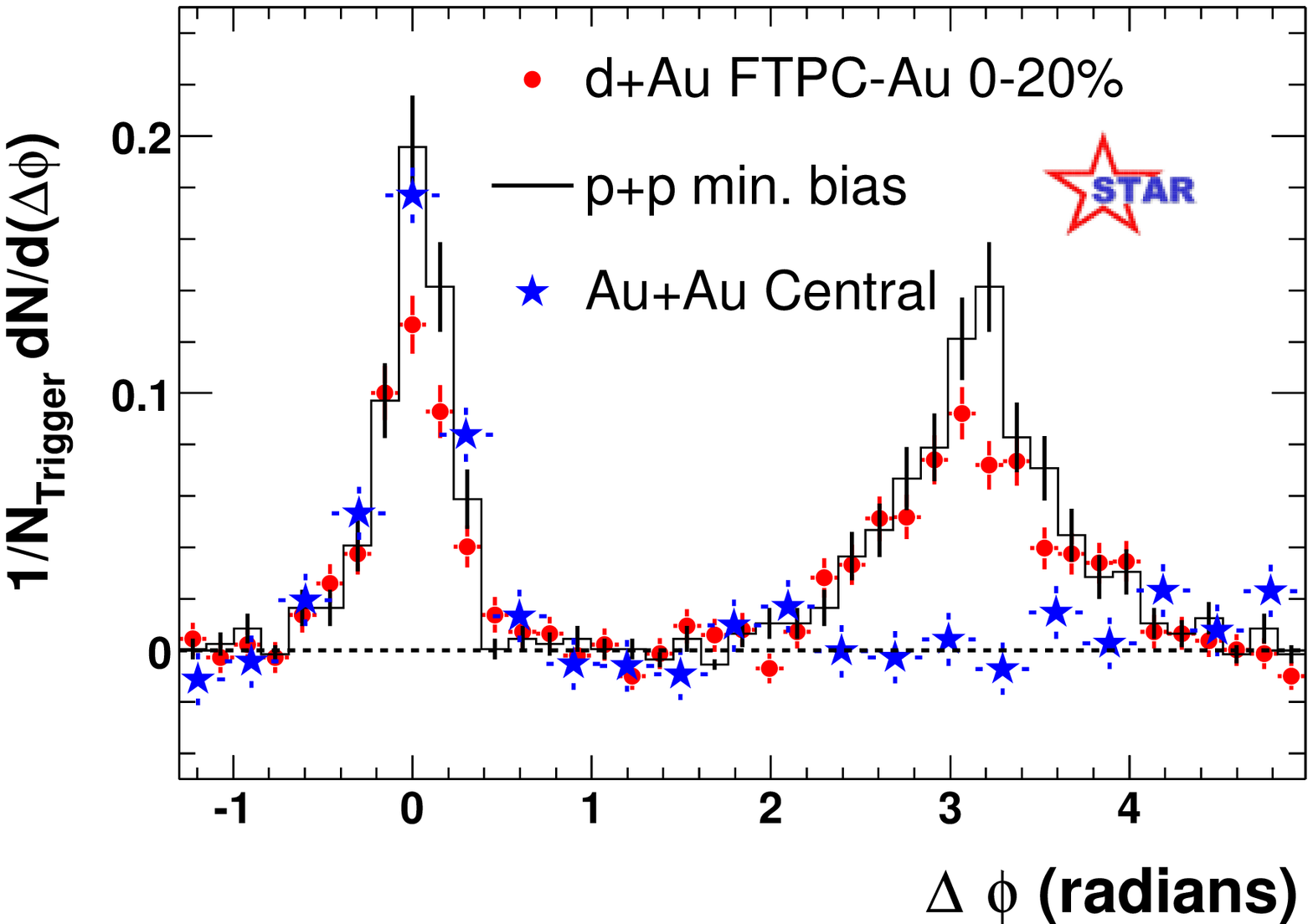}
\end{minipage}
\begin{minipage}{0.5\textwidth}
\center
\includegraphics[height=51mm,width=81mm]{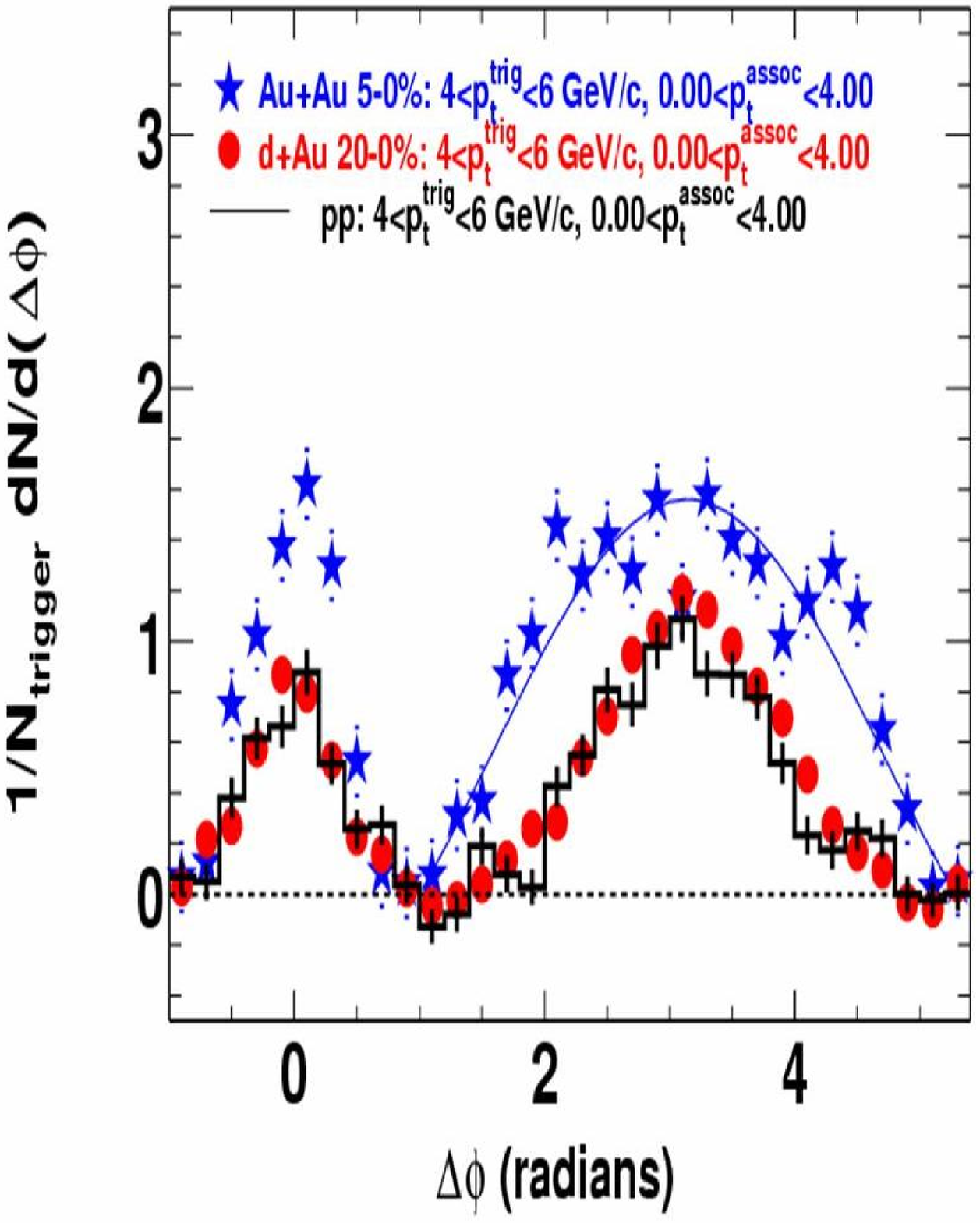}
\end{minipage}
\caption{(a) STAR data on dihadron angular correlations. $\Delta \phi$
  is the opening angle between the trigger ($4 < p_{T}^{trig} < 6$
  GeV$/c$) and associated particles ($2 < p_T^{assoc} < p_T^{trig}$
  GeV$/c$). Figure from \cite{Adams:2003im}. (b) Similar to the left
  panel, except that $0 < p_T^{assoc} < 4$ GeV$/c$. Figure from
  \cite{Adams:2005ph}.}
\label{jq2}
\end{figure}

Figure \ref{jq1}(b) illustrated jet quenching in a single-particle
inclusive yield. Jet quenching is also seen in dihadron angular
correlations shown in Fig. \ref{jq2} as a function of the opening
angle between the trigger and associated particles. The only
difference between the left and the right panels is the definition of the
associated particles. The left panel shows the suppression of the
away-side jet in AuAu central, but not in $pp$ and $d$Au central
collisions. This is expected because unlike AuAu collisions, no hot
and dense medium is likely to be formed in $pp$ and $d$Au collisions,
and so there is no quenching of the away-side jet. Energy of the
away-side parton in a AuAu collision is dissipated in the medium
thereby producing low-$p_T$ or soft particles. When even the soft
particles are included, the away-side jet reappears in the AuAu data
as shown in the right panel. Its shape is broadened due to
interactions with the medium.

\begin{figure}[htbp]
\begin{minipage}{0.5\textwidth}
\center \includegraphics[scale=0.8]{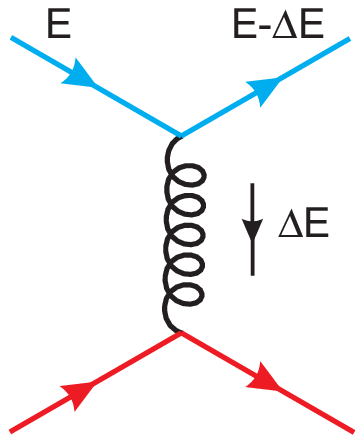}
\end{minipage}
\begin{minipage}{0.5\textwidth}
\center
\includegraphics[scale=0.8]{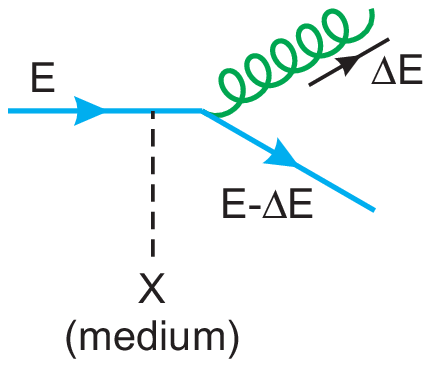}
\end{minipage}
\caption[]{Collisional (left) and medium-induced radiative (right)
  energy loss mechanisms. Their predictions for the energy loss per 
unit length differ from each other: $\Delta E \propto L$ and 
$\Delta E \propto L^2$, respectively.
 Figure from \cite{d'Enterria:2009am}.}
\label{Eloss}
\end{figure}

Figure \ref{Eloss} shows two main mechanisms by which a parton moving
in the medium loses energy. Collisional energy loss via elastic
scatterings dominates at low momenta whereas the radiative energy loss
via inelastic scatterings dominates at high momenta. Energy loss per
unit path length depends on the properties of the parton (parton
species, energy $E$), as well as the properties of the medium
($T,~dN^g/dy,~\hat{q}$). The jet quenching parameter, $\hat q$, is
defined as the average $p_T^2$ transferred to the outgoing parton per
unit path length. The value of $\hat q$ estimated in leading-order QCD
is $\simeq 2.2$ GeV$^2$/fm, while the value extracted from
phenomenological fits to the RHIC experimental data on parton energy
loss is $\mathcal{O}(10)$ GeV$^2$/fm.

\begin{figure}[htbp]
\centering\includegraphics[width=0.9\linewidth]{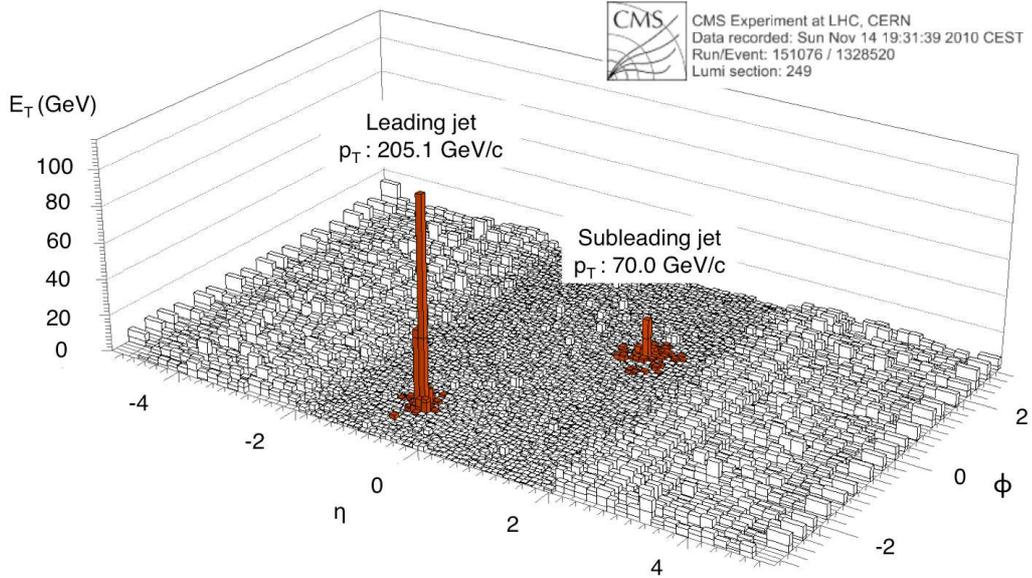}
\caption{Jet quenching in PbPb collision at the centre-of-mass
  energy $\sqrt{s_{NN}}=2.76$ TeV at CMS. $E_T$ is the summed
  transverse energy in the electromagnetic and hadron
  calorimeters. $\eta$ and $\phi$ are the pseudorapidity and azimuthal
  angle, respectively. Figure from \cite{Chatrchyan:2011sx}.}
\label{jetlhc}
\end{figure}

Jets are more abundant and easier to reconstruct at LHC than at RHIC. Figure
\ref{jetlhc} shows an example of an unbalanced dijet in a PbPb
collision event at CMS (LHC). By studying the evolution of the dijet
imbalance as a function of collision centrality and energy of the
leading jet, one hopes to get an insight into the dynamics of the jet
quenching.

For recent reviews of jet quenching, see \eg
\cite{Wiedemann:2009sh,d'Enterria:2009am,oai:arXiv.org:1002.2206,
Spousta:2013aaa}.

% -------------------------------------------------------------------------

\section{Some other important observables}

Elliptic flow, or more generally anisotropic collective flow, and jet
quenching, which we discussed above are examples of soft and hard
probes, respectively. Here `soft' refers to the low-$p_T$ regime: $0
\lsim p_T \lsim 1.5$ GeV$/c$, and hard refers to high-$p_T$ regime: $p_T
\gg 5$ GeV$/c$. (At RHIC, such high-$p_T$ jets are rare, which 
explains the relatively low $p_T$ cuts used in Fig. \ref{jq2}.) 
The medium-$p_T$ regime ($1.5 \lsim p_T \lsim 5$ GeV$/c$)
is also interesting, \eg for the phenomenon of constituent quark
number scaling or quark coalescence. In this section we discuss
briefly this and other important observables. We shall, however, not
discuss a few other important topics such as femtoscopy with
two-particle correlation measurements
\cite{Tomasik:2002rx,Padula:2004ba,Lisa:2005dd} and electromagnetic
probes of QGP \cite{David:2006sr,Tserruya:2009zt}.

\subsection{Constituent quark number scaling}

\begin{figure}[htbp]
\centering\includegraphics[width=0.9\linewidth]{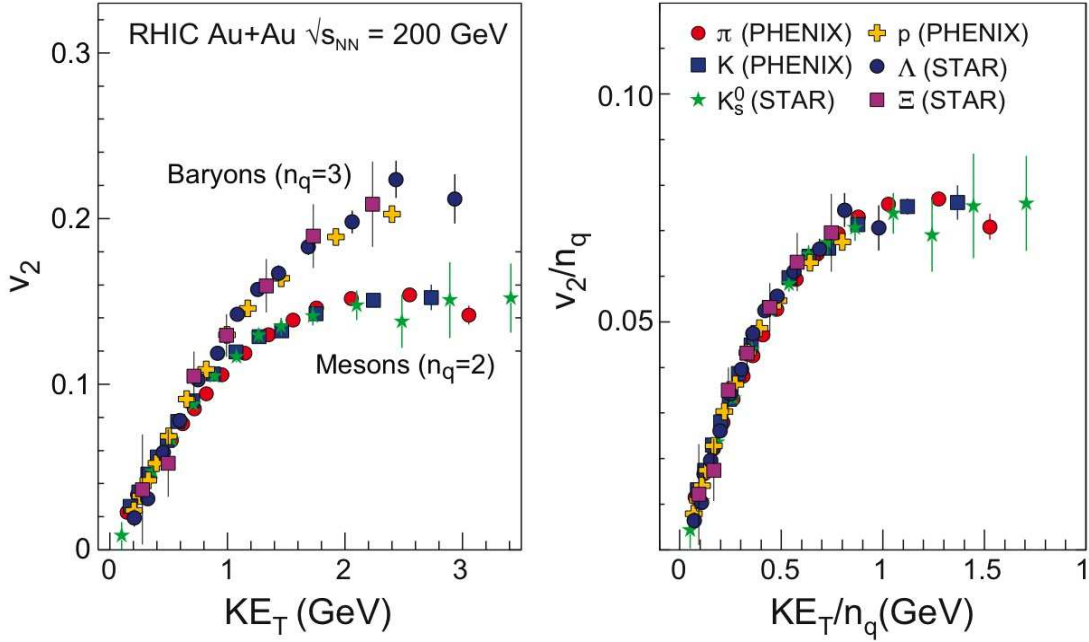}
\caption{(Left) Elliptic flow $v_2$ vs transverse kinetic energy
  $KE_T$ for various baryons and mesons. (Right) Both $v_2$ and $KE_T$
  are scaled by the number of constituent quarks $n_q$. Figure from
  \cite{Muller:2007rs}.}
\label{cqs}
\end{figure}

In the high-$p_T$ regime, hadronization occurs by fragmentation,
whereas in the medium-$p_T$ regime, it is modelled by quark
recombination or coalescence. The phenomenon of constituent quark
number scaling provides experimental support to this model. Figure
\ref{cqs} explains the meaning of constituent quark number ($n_q$) scaling. In
the left panel one sees two distinct branches, one for baryons
($n_q=3$) and the other for mesons ($n_q=2$). When scaled by $n_q$
(right panel), the two curves merge into one universal curve,
suggesting that the flow is developed at the quark level, and hadrons
form by the merging of constituent quarks. This observation provides the
most direct evidence for deconfinement so far. 
ALICE (LHC) has also reported results for the elliptic flow $v_2(p_T)$ of
identified particles produced in PbPb collisions at 2.76 TeV. The
constituent quark number scaling was found to be not as good as at
RHIC \cite{oai:arXiv.org:1212.1292}.

For a recent review see \cite{Fries:2008hs}.

% -------------------------------------------------------------------------

\subsection{Ratios of particle abundances\footnote{See also section 6.2.2.}}

Ratios of particle abundances such as $K/\pi,~ p/\pi$, etc. constrain
models of particle production. In the thermal or statistical 
hadronization model
\cite{BraunMunzinger:2003zd,Becattini:2009sc}, 
particles in the final state are assumed
to be emitted by a source in a thermodynamic equilibrium characterized
by only a few parameters such as the (chemical freezeout) temperature
and the baryo-chemical potential. These parameters are determined by
fitting the experimental data on particle abundances. This model has
been quite successful in explaining the Alternating Gradient
Synchrotron (AGS), Super Proton Synchrotron (SPS), and RHIC data on the
particle ratios \cite{Cleymans:1998fq,Andronic:2008gu}. These
facilities together cover the centre-of-mass energy ($\sqrt{s_{NN}}$)
range from 2 GeV to 200 GeV.

For a recent review of the statistical hadronization picture with an
emphasis on charmonium production, see \cite{BraunMunzinger:2009ih}.

% -------------------------------------------------------------------------

\subsection{Strangeness enhancement}

\begin{figure}[htbp]
\centering\includegraphics[scale=0.6]{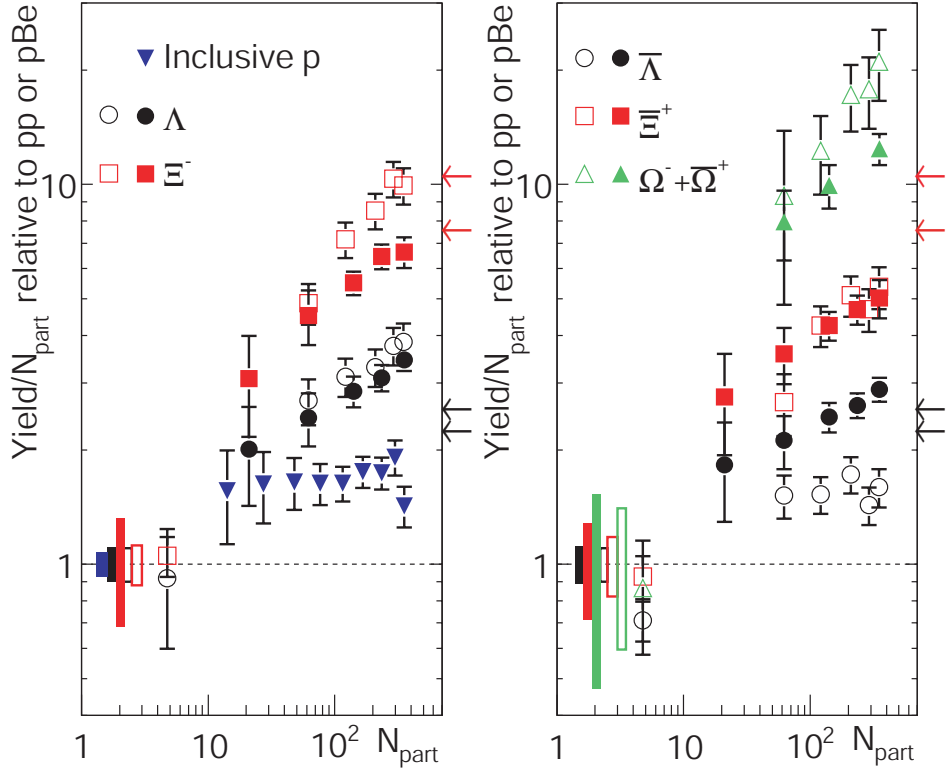}
\caption{Enhanced strange baryon production as a function of
  $\mean{N_{part}}$, at mid-rapidity, in $AA$ collisions compared to
  $\mean{N_{part}}$-scaled $pp$ interactions at the same energy. Solid
  markers: STAR data on AuAu collisions at $\sqrt{s_{NN}}=200$
  GeV. Open symbols: SPS data on PbPb collisions at
  $\sqrt{s_{NN}}=17.3$ GeV. Boxes at unity show statistical and
  systematic uncertainties and arrows on the right axes mark the
  predictions of a thermal model. Figure from \cite{Abelev:2007xp}.}
\label{strenh}
\end{figure}

Production of strange particles is expected to be enhanced
\cite{Rafelski:1982pu,Koch:1986ud} in relativistic nucleus-nucleus
collisions relative to the scaled up $pp$ data (Eq. (\ref{yield}))
because of the following reasons: (1) Although $m_s \gg m_{u,d}$,
strange quarks and antiquarks can be abundant in an equilibrated QGP
with temperature $T > m_s$, (2) large gluon density in QGP leads to an
efficient production of strangeness via gluon fusion $gg \rightarrow s
{\bar s}$, and (3) energy threshold for strangeness production
in the purely hadron-gas scenario is much higher than in QGP. Abundance of
strange quarks and antiquarks in QGP is expected to leave its imprint
on the number of strange and multi-strange hadrons detected in the
final state. The above expectation was borne out by the measurements
made at SPS and RHIC; see Fig. \ref{strenh} where $N_{part}$ is the
mean number of participating nucleons in a nucleus-nucleus collision,
estimated using the Glauber model \cite{Miller:2007ri} and serves as a
measure of the centrality of the collision. The idea of strangeness
enhancement in $AA$ collisions or equivalently of strangeness
suppression in $pp$ collisions can be recast in the language of
statistical mechanics of grand canonical (for central $AA$ collisions)
and canonical (for $pp$ collisions) ensembles; see, e.g.,
\cite{Abelev:2007xp}. A complete theoretical understanding of these
results is yet to be achieved \cite{Abelev:2007xp}.

For a review of strange hadron production in heavy-ion collisions from
SPS to RHIC, see \cite{Blume:2011sb}. For the ALICE (LHC) results on
multi-strange baryon production at 2.76 TeV, see \cite{ABELEV:2013zaa}.
ALICE observed that the strangeness enhancement was less pronounced than
at lower energies.

% -------------------------------------------------------------------------

\subsection{Sequential melting of heavy quarkonia\footnote{See also 
section 6.2.3.}}

Colour Debye screening of the attraction between heavy quarks ($c$ or
$b$) and antiquarks ($\bar c$ or $\bar b$) in a hot and dense medium
such as QGP is expected to suppress the formation of quarkonia
relative to what one expects from a $pp$ baseline measurement
\cite{Matsui:1986dk}. Observation of suppression would thus serve as a
signal for deconfinement. As the temperature of the medium rises,
various quarkonium states are expected to `melt' one by one 
in the sequence of their increasing binding energies.
The sequential melting of heavy quarkonia thus serves as a `thermometer'
for the medium. A reliable estimation of the
charmonium\footnote{Similar statements would be true for the
  bottomonium.} formation rates, however, needs to take into account
several other competing effects:
\begin{itemize}
\item
gluon shadowing/anti-shadowing and saturation effects in the initial
wave functions of the colliding nuclei,
\item
initial- and final-state $k_T$ scatterings and parton-energy loss,
\item
charmonium formation via colour-singlet and colour-octet channels,
\item
feed-down from the excited states of the
charmonium to its ground state,
\item
secondary charmonium production by
recombination or coalescence of independently produced $c$ and $\bar
c$,
\item
interaction of the outgoing charmonium with the medium, etc.
\end{itemize}
A systematic study of suppression patterns of $J/\psi$ and $\Upsilon$
families, together with $pA$ baseline measurements, over a broad
energy range, would help disentangle these hot and cold nuclear matter
effects.

For reviews of charmonium and/or bottomonium production in heavy-ion
collisions, see \cite{Kluberg:2009wc,Linnyk:2008hp,Rapp:2008tf,
Brambilla:2010cs}. For
a review of heavy-flavour probes of the QCD matter formed at RHIC, see
\cite{Frawley:2008kk}.

% -------------------------------------------------------------------------

\section{Big Bang and Little Bang}

Having described the various stages in the relativistic heavy-ion
collisions and the most important observables and probes in this
field, let me bring out the striking similarities between the Big Bang
and the Little Bang. In both cosmology and the physics of relativistic
heavy-ion collisions, the initial quantum fluctuations ultimately lead
to macroscopic fluctuations and anisotropies in the final state. In
both the fields, the goal is to learn about the early state of the
matter from the final-state observations. See Table 1 for the
comparison of these two fields. Here 2.73 K and 1.95 K are photon and
neutrino decoupling or freezeout temperatures, respectively. $T_{ch}$
and $T_{kin}$ are the chemical and kinetic freezeout temperatures
mentioned in section 1. The last two rows list the various
experimental `tools' and the years in which they were
commissioned. For a more detailed comparison, see
\cite{yagi,Mishra:2007tw,Mishra:2008dm}.

\begin{table}
\caption{Big Bang and Little Bang comparison}
\begin{center}
\begin{tabular}{|c|c|c|} \hline
 & Big Bang & Little Bang \\ \hline\hline
Occurrence & Only once & Millions of times at RHIC, LHC\\ \hline
Initial state & Inflation? ($10^{-35}$ s) & Glasma? ($10^{-24}$ s) \\ \hline 
Expansion & General Relativity & Rel. imperfect fluid dynamics\\ \hline
Freezeout temperatures & $\gamma: 2.73$ K, $\nu: 1.95$ K & 
$T_{ch}\sim 150, T_{kin}\sim 120$ MeV \\ \hline
Anisotropy in & Final temp. (CMB) & Final flow profile \\ \hline
Penetrating probes & Photons & Photons, jets \\ \hline
Chemical probes & Light nuclei & Various hadron species \\ \hline
Colour shift & Red shift & Blue shift \\ \hline
Tools & COBE, WMAP, Planck & SPS, RHIC, LHC \\ \hline
Starting years & 1989, 2001, 2009 & 1987, 2000, 2009 \\ \hline
\end{tabular}
\end{center}
\end{table}

% -------------------------------------------------------------------------

\section{Fluid dynamics}

The kinetic or transport theory of gases is a microscopic description
in the sense that detailed knowledge of the motion of the constituents
is required. Fluid dynamics (also loosely called hydrodynamics) is an
effective (macroscopic) theory that describes the slow,
long-wavelength motion of a fluid close to local thermal
equilibrium. No knowledge of the motion of the constituents is
required to describe observable phenomena. Quantitatively, if $l$
denotes the mean free path, $\tau$ the mean free time, $k$ the wave
number, and $\omega$ the frequency, then $kl \ll 1,~ \omega \tau \ll
1$ is the hydrodynamic regime, $kl \simeq 1,~ \omega \tau \simeq 1$
the kinetic regime, and $kl \gg 1,~ \omega \tau \gg 1$ the
nearly-free-particle regime.

Relativistic hydrodynamic equations are a set of coupled partial
differential equations for number density $n$, energy density
$\epsilon$, pressure $P$, hydrodynamic four-velocity $u^\mu$, and
in the case of imperfect hydrodynamics, also
bulk viscous pressure $\Pi$, particle-diffusion
current $n^\mu$, and shear stress tensor $\pi^{\mu\nu}$. In addition,
these equations also contain the coefficients of shear and bulk
viscosities and thermal conductivity, and the corresponding relaxation
times. Further, the equation of state (EoS) needs to be supplied to
make the set of equations complete. Hydrodynamics is a powerful
technique: Given the initial conditions and the EoS, it predicts the
evolution of the matter. Its limitation is that it is applicable at or
near (local) thermal equilibrium only.

Relativistic hydrodynamics finds applications in cosmology, astrophysics,
high-energy nuclear physics, etc. In relativistic heavy-ion
collisions, it is used to calculate the multiplicity and transverse
momentum spectra of hadrons, anisotropic flows, and femtoscopic
radii. Energy density or temperature profiles resulting from the
hydrodynamic evolution are needed in the calculations of jet
quenching, $J/\psi$ melting, thermal photon and dilepton productions,
etc. Thus hydrodynamics plays a central role in modeling relativistic
heavy-ion collisions.

\begin{figure}[htbp]
\centering\includegraphics[width=0.5\linewidth]{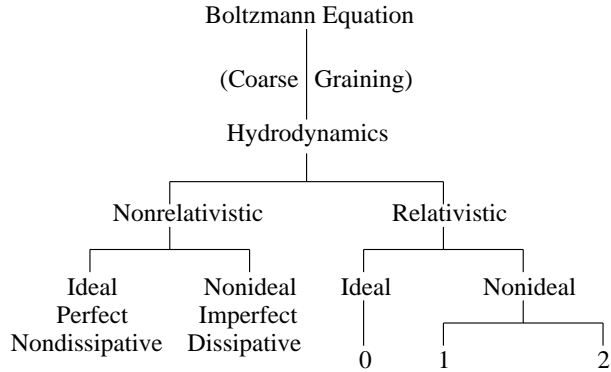}
\caption{Coarse-Graining of the Boltzmann equation}
\label{cgkt}
\end{figure}

Hydrodynamics is formulated as an order-by-order expansion
in the sense that in the first (second)-order
theory, the equations for the dissipative fluxes contain the first
(second) derivatives of $u^\mu$.
The ideal hydrodynamics is called the zeroth-order theory. 
The zeroth-, first-, and second-order equations are named after
Euler, Navier-Stokes, and Burnett, respectively, in the
non-relativistic case (Fig. \ref{cgkt}). The relativistic
Navier-Stokes equations are parabolic in nature and exhibit acausal
behaviour, which was rectified in the (relativistic 
second-order) Israel-Stewart
(IS) theory \cite{Israel:1979wp}. The formulation of the relativistic
imperfect second-order hydrodynamics (`2' in Fig. \ref{cgkt}) is
currently under intense investigation; see, e.g.,
\cite{Denicol:2012cn,Denicol:2012es,Jaiswal:2013fc,Jaiswal:2012qm,
Bhalerao:2013pza}
for the recent activity in this area.
Hydrodynamics has traditionally been derived either from entropy
considerations (i.e., the generalized second law of thermodynamics) or
by taking the second moment of the Boltzmann equation.

For a comprehensive treatment of relativistic hydrodynamics, numerical
techniques, and applications, see \cite{rezzolla}. For an elementary
introduction to relativistic hydrodynamics and its application to
heavy-ion collisions, see \cite{Ollitrault:2007du}. For a review of
new developments in relativistic viscous hydrodynamics, see
\cite{Romatschke:2009im}.

% ----------------------------------------------------------------------------

\section{LHC highlights}

\subsection{RHIC-LHC comparison}

Table 2 compares some basic results obtained at LHC soon after it
started operating, with similar results obtained earlier at RHIC.
Here $dN_{ch}/d\eta$ is the charged particle pseudorapidity 
density, at mid-rapidity, 
normalized by $\mean{N_{part}}/2$ where $\mean{N_{part}}$ is the mean
number of participating nucleons in a nucleus-nucleus collision,
estimated using the Glauber model \cite{Miller:2007ri}.
$\epsilon_{Bj}$ is the initial energy density estimated using the
well-known Bjorken formula \cite{yagi,wong}.  $\tau_i$ is the initial
or formation time of QGP. Assuming conservatively the same 
$\tau_i \simeq 0.5$ fm at
LHC as at RHIC, one gets an estimate of $\epsilon_{Bj}$ at LHC. $T_i$
is the initial temperature fitted to reproduce the observed
multiplicity of charged particles in a hydrodynamical model. 
Note that the $\sim 30$\% increase in $T_i$ is consistent with the
factor of $\sim 3$ rise in $\epsilon_{Bj}$.
$V_{f.o.}$
is the volume of the system at the freezeout, measured with two-pion
Bose-Einstein correlations. $v_{flow}$ is the radial
velocity of the collective flow of matter. $v_2$ is the elliptic
flow. It is clear from Table 2 that the QGP fireball produced at LHC
is hotter, larger, and longer-lasting, as compared with that at RHIC.
\begin{table}
\caption{RHIC-LHC comparison}
\begin{center}
\begin{tabular}{|c|c|c|c|} \hline
& RHIC (AuAu) & LHC (PbPb) & Increase by factor or \%\\ \hline
$\sqrt {s_{NN}}$ (GeV) & 200 & 2760 & 14 \\ \hline
$dN_{ch}/d \eta/\left( \frac{< N_{part} >}{2} \right)$ & 3.76 & 8.4 & 2.2 \\ \hline
$\epsilon_{Bj} \tau_i$ (GeV/fm$^2$) & 16/3 & 16 & 3 \\ \hline
$\epsilon_{Bj}$ (GeV/fm$^3$) & 10 & 30 & 3 \\ \hline
$T_i$ (MeV) & 360 & 470 & 30\% \\ \hline
$V_{f.o.}$ (fm$^3$) & 2500 & 5000 & 2 \\ \hline
Lifetime (fm$/c$) & 8.4 & 10.6 & 26\% \\ \hline
$v_{flow}$ & 0.6 & 0.66 & 10\% \\ \hline
$<p_T>_\pi$ (GeV) & 0.36 & 0.45 & 25\% \\ \hline
Differential $v_2(p_T)$ & & & unchanged \\ \hline
$p_T$-integrated $v_2$ & & & 30\% \\ \hline
\end{tabular}
\end{center}
\end{table}

% ----------------------------------------------------------------------------

\subsection{Some surprises at LHC}

\subsubsection{Charged-particle production at LHC}

\begin{figure}[htbp]
\begin{minipage}{0.5\textwidth}
\center
\includegraphics[scale=0.40]{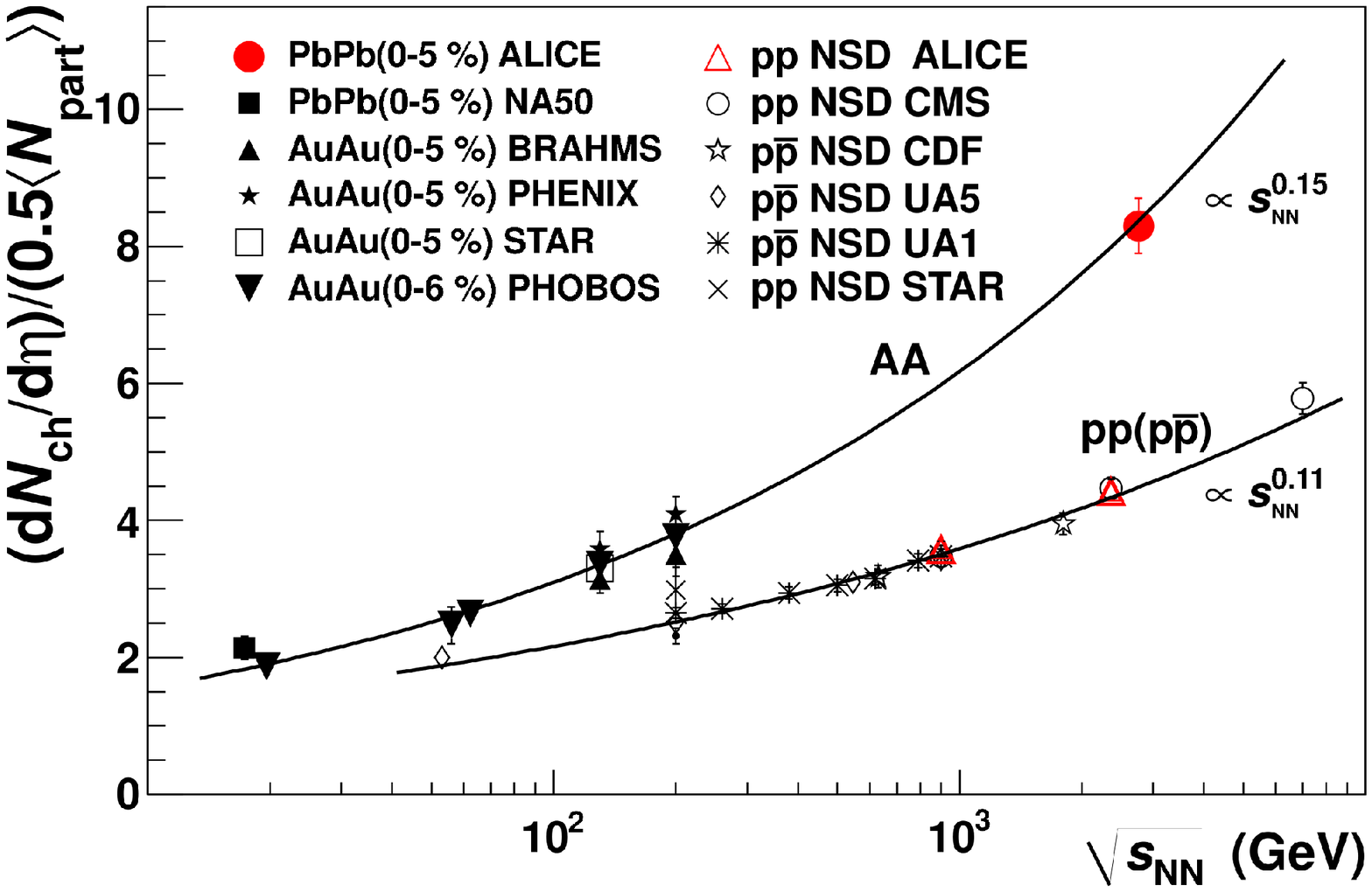}
\end{minipage}
\begin{minipage}{0.5\textwidth}
\center
\includegraphics[scale=0.40]{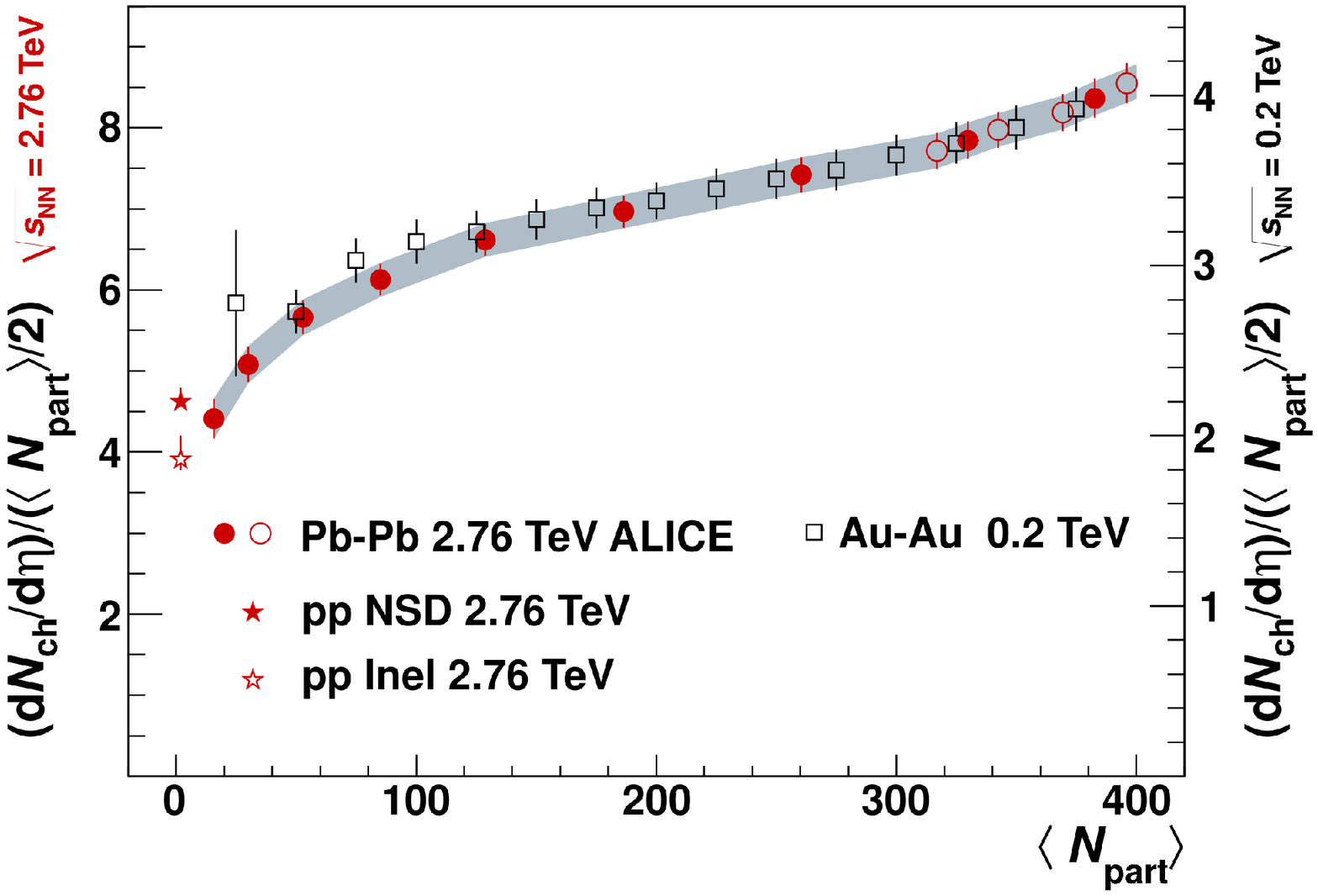}
\end{minipage}
\caption{Charged-particle pseudorapidity density (at $\eta=0$) per
  colliding nucleon pair vs $\sqrt{s_{NN}}$ (left panel) and
  $\mean{N_{part}}$ (right panel). Figure from \cite{Reygers:2012yc}.}
\label{cpp}
\end{figure}

Figure \ref{cpp} presents perhaps the most basic observable in
heavy-ion collisions --- the number of charged particles produced.
This observable helps place constraints on the particle production
mechanisms and provides a first rough estimate of the initial energy
density reached in the collision. The left panel compares the
charged-particle production in central $AA$ and non-single-diffractive
(NSD)\footnote{Non-single-diffractive $pp$ collisions are those which
  exclude the elastic scattering and single-diffractive events.} $pp
(p \bar p)$ collisions at various energies and facilities. The curves
are simple parametric fits to the data; note the higher power of
$s_{NN}$ in the former case. The precise magnitude of $dN_{ch}/d\eta$
measured in PbPb collisions at LHC was somewhat on a higher side
than expected. Indeed, as is clear from the figure, the logarithmic
extrapolation of the lower-energy measurements at AGS, SPS, and RHIC
grossly under-predicts the LHC data. The right panel highlights an even
more surprising fact that the shape of the plotted observable vs
centrality is nearly independent of the centre-of-mass energy, except
perhaps for the most peripheral $AA$ collisions. Studying the centrality
dependence of the charged-particle production throws light on the roles
played by hard scatterings and soft processes. For details, see 
\cite{Abbas:2013bpa}.

% ----------------------------------------------------------------------------
\subsubsection{Particle ratios at LHC --- Proton anomaly}

\begin{figure}[htbp]
\begin{minipage}{0.50\textwidth}
\center
\includegraphics[scale=0.3528]{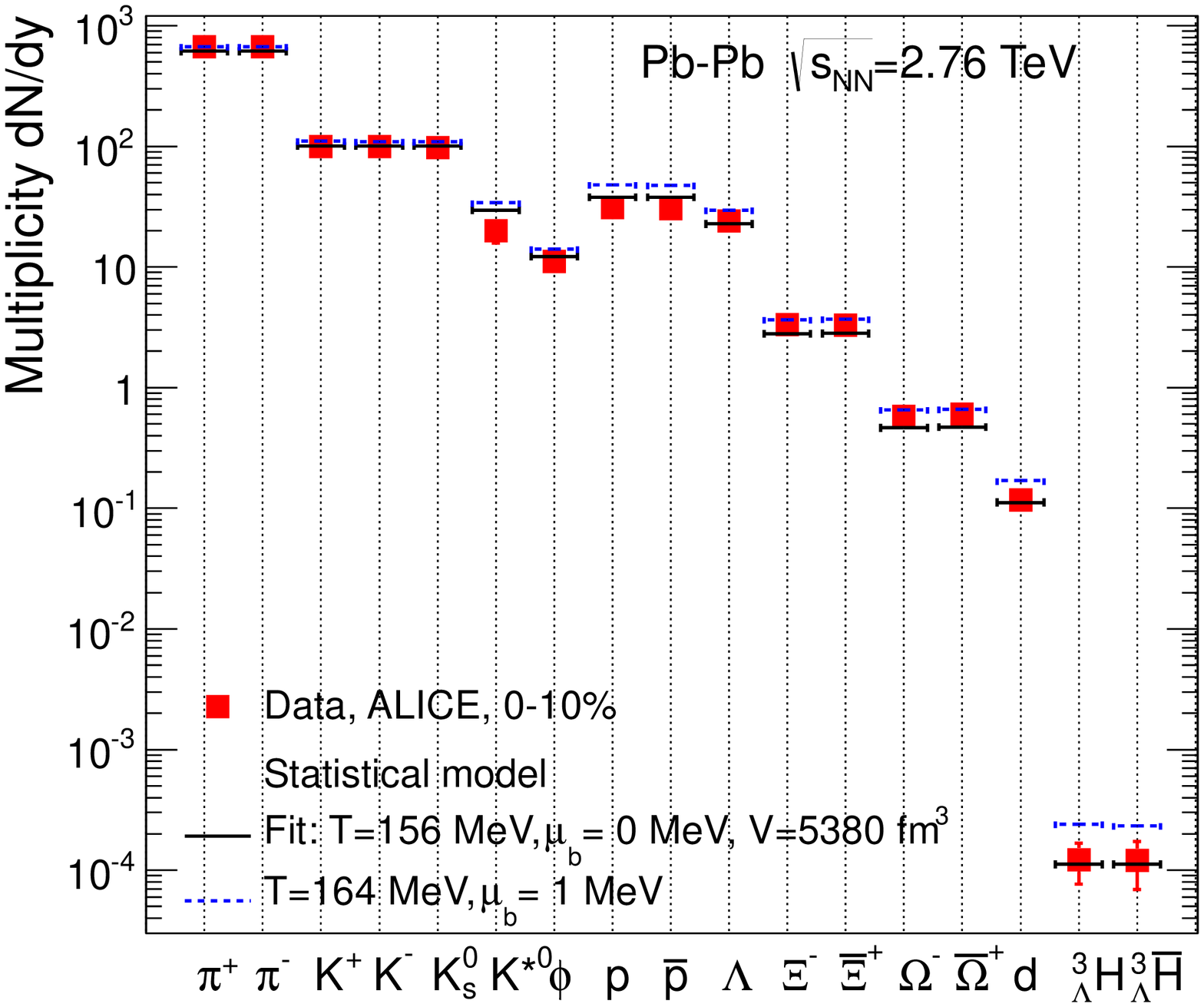}
\end{minipage}
\begin{minipage}{0.50\textwidth}
\center
\includegraphics[scale=0.3969]{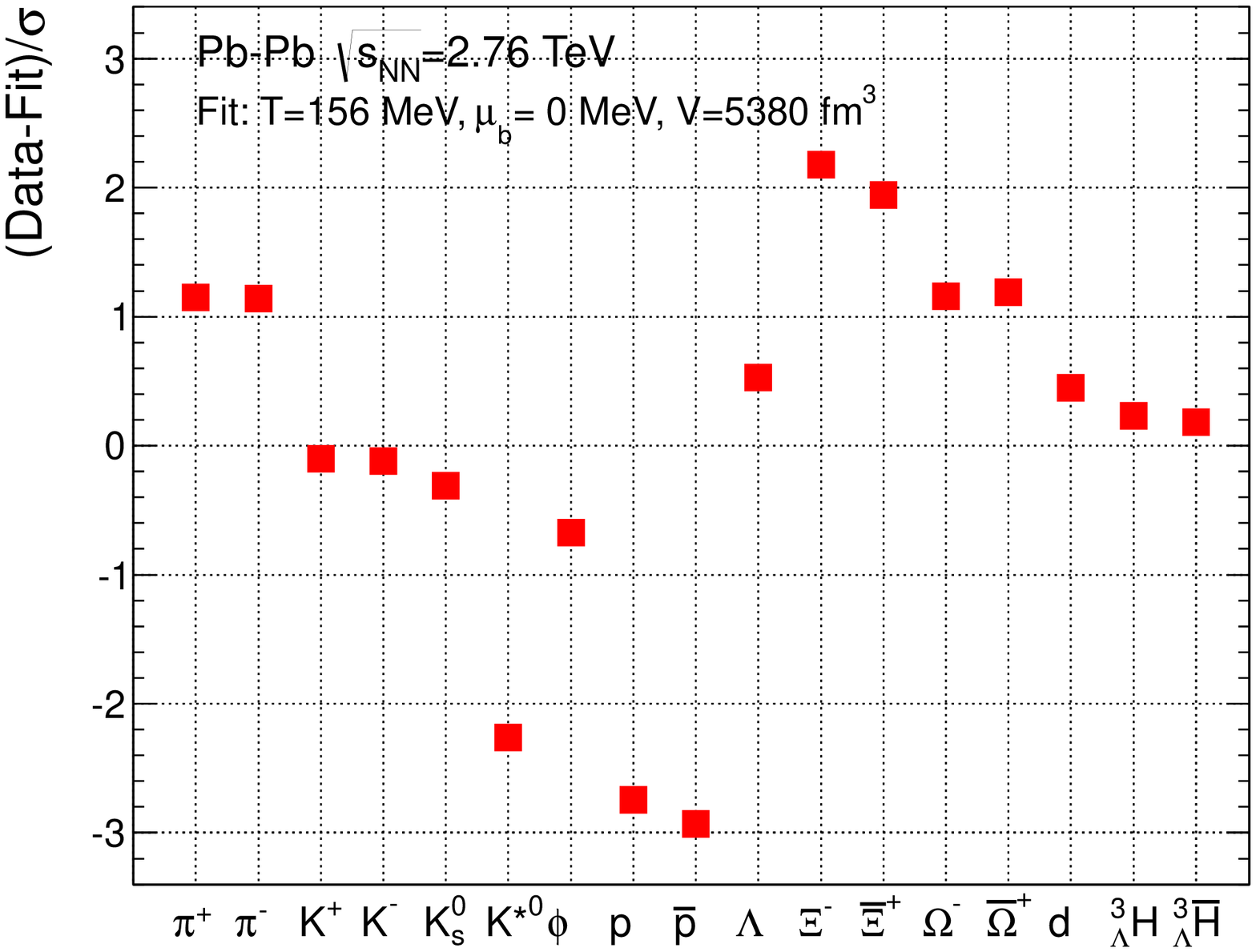}
\end{minipage}
\caption{Left: Hadron yields from ALICE (LHC) together with the fit
  based on the thermal model (solid black lines). The data point for
  $K^{0*}$ is not included in the fit. Blue dotted lines show results
  of the model for the indicated values of $T$ and $\mu_b$, normalized
  to the value for $\pi^+$. Right: Deviations between the thermal fit
  and the data.  Note that the $p$ and $\bar{p}$ yields are below the
  thermal fit by 2.7 and 2.9 sigma, respectively, whereas the cascade yields
  are above the fit by about two sigma. Figures from
  \cite{Stachel:2013zma}.}
\label{ratios}
\end{figure}

We described above in section 3.2 the success of the
thermal/statistical hadronization model in explaining the ratios of
particle abundances measured at AGS, SPS, and RHIC. When extended to
the LHC energies, however, the model was unable to reproduce the
$p/\pi^+$ and $\bar{p}/\pi^-$ ratios; the absolute $p,\bar{p}$ yields
were off by almost three standard deviations (Fig. \ref{ratios}).
Current attempts to understand these discrepancies focus on the
possible effects of (a) as yet undiscovered hadrons, or in other
words, the incomplete hadron spectrum, (b) the annihilation of some
$p,\bar{p}$ in the final hadronic phase, or (c) the out-of-equilibrium
physics currently missing in the model. None of these effects has been
found to be satisfactory because while reducing the (Data-Fit)
discrepancy at one place, it worsens it at other place(s)
\cite{Stachel:2013zma}. Finally, Fig. \ref{ratios} also shows that
most antiparticle/particle ratios are unity within error bars
indicating a vanishing baryo-chemical potential at LHC.

% ----------------------------------------------------------------------------

\subsubsection{Quarkonium story at LHC}

\begin{figure}[htbp]
\begin{minipage}{0.5\textwidth}
\center
\includegraphics[scale=0.40]{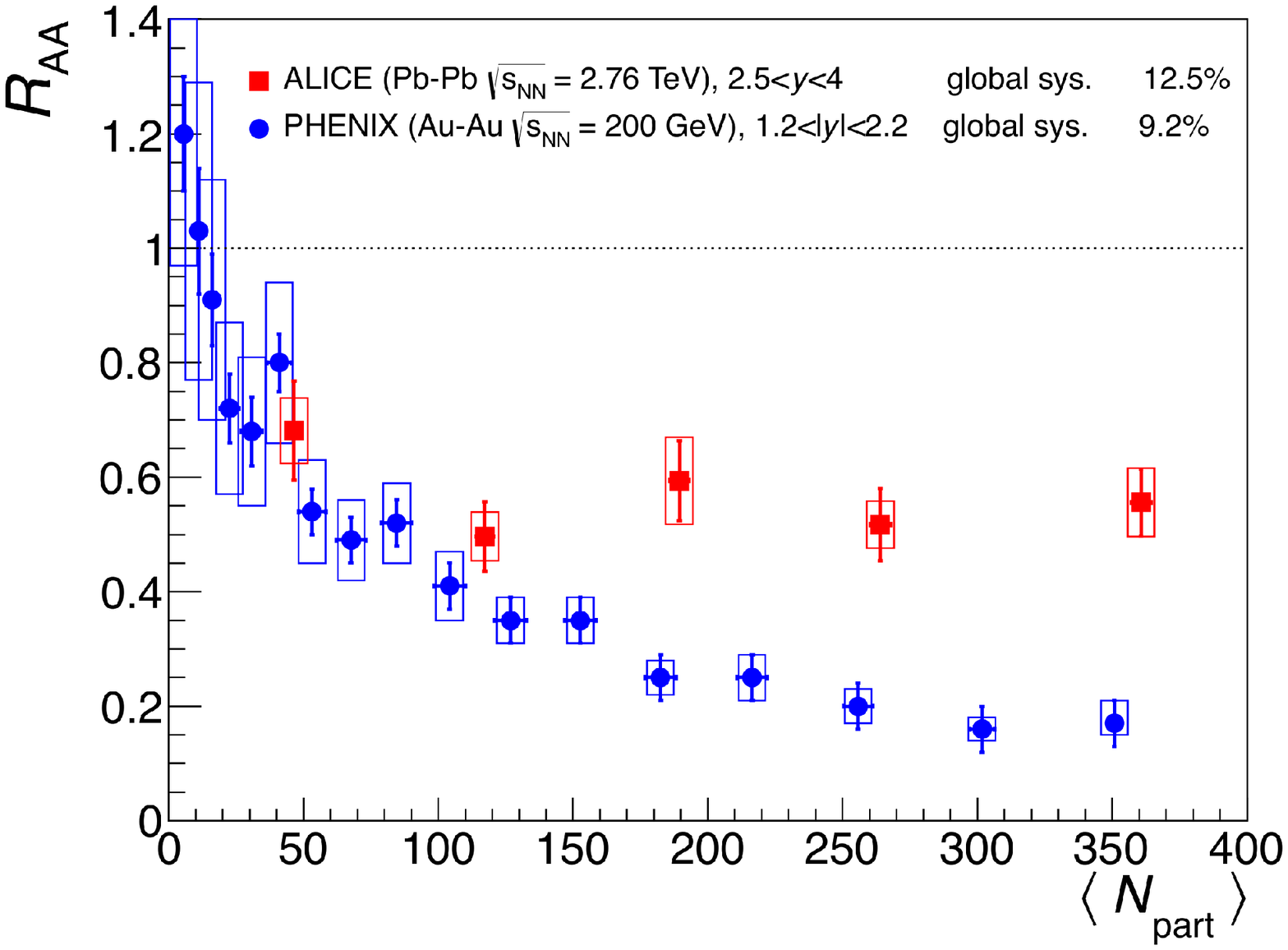}
\end{minipage}
\begin{minipage}{0.49\textwidth}
\center
\includegraphics[scale=0.63]{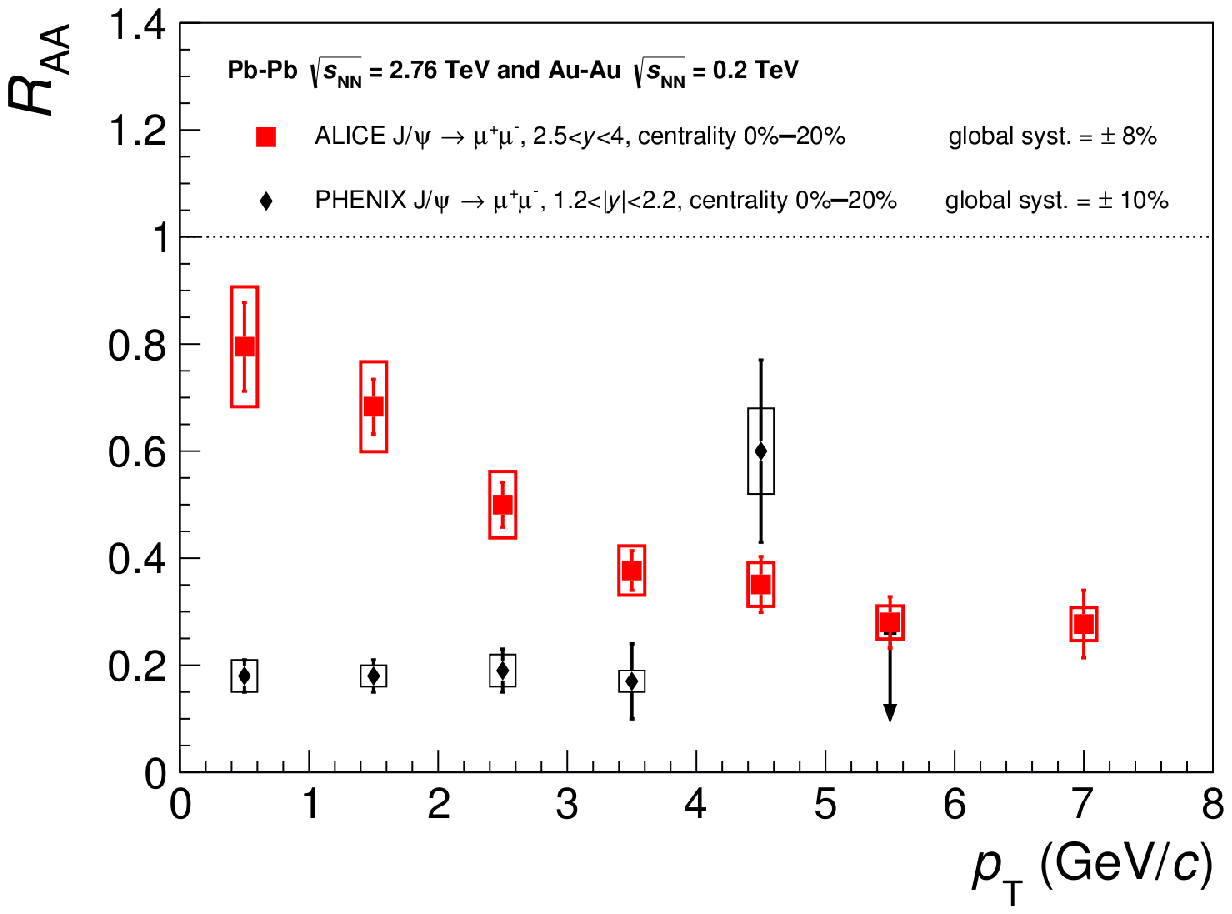}
\end{minipage}
\caption{Nuclear modification factor $R_{AA}$ of $J/\psi$ vs
  centrality (left) and $p_T$ (right). Figures from
  \cite{Reygers:2012yc} and \cite{Abelev:2013ila}.}
\label{RAAJ}
\end{figure}

We described above in section 3.4 the melting of heavy quarkonium as a
possible signature of deconfinement or colour screening effects in
QGP. Anomalous suppression of $J/\psi$ was first seen at SPS. No
significant differences in the suppression pattern were observed at
RHIC. LHC, however, has thrown some surprises which are not yet fully
understood.  Figure \ref{RAAJ} presents the nuclear modification
factor $R_{AA}$ of $J/\psi$ as a function of centrality (left) and
$p_T$ (right), at similar rapidities.  Note the differences between
the PHENIX and ALICE measurements. Differences at low $p_T$ in the
right-hand panel are possibly because of the larger recombination
probability at ALICE than at PHENIX; this probability is expected to
decrease at high $p_T$. Sequential suppression of upsilon states was
observed by CMS in PbPb collisions at 2.76 TeV: The $R_{AA}$ values
for $\Upsilon$(1S), $\Upsilon$(2S), and $\Upsilon$(3S), were about
0.56, 0.12, and lower than 0.10, respectively \cite{upsilon}. For the
status of the evolving quarkonium saga, see \cite{Tserruya:2013rca}.

% ----------------------------------------------------------------------------

\section{Concluding remarks}

(1) Quark-gluon plasma has been discovered, and we are in the midst of
trying to determine its thermodynamic and transport properties accurately.

\noindent (2) Data on the collective flow at RHIC/LHC have provided a
strong support to hydrodynamics as the appropriate effective theory
for relativistic heavy-ion collisions. The most complete event-to-event
hydrodynamic calculations to date \cite{Gale:2012rq,Gale:2013da} have
yielded $\eta/s=0.12$ and 0.20 at RHIC (AuAu, 200 GeV) and LHC
(PbPb, 2.76 TeV), respectively, with at least 50\% systematic
uncertainties. These are the average values over the temperature
histories of the collisions. Uncertainties associated with (mainly)
the initial conditions have so far prevented a more precise
determination of $\eta/s$.

\noindent (3) Surprisingly, even the $pp$ collision data at 7 TeV are
consistent with the hydrodynamic picture, if the final multiplicity is
sufficiently large!

\noindent (4) An important open question is at what kinematic scale
partons lose their quasiparticle nature (evident in jet quenching) and
become fluid like (as seen in the collective flow)?

\noindent (5) QCD phase diagram still remains largely unknown.

\noindent (6) RHIC remains operational. ALICE, ATLAS, and CMS at 
LHC all have come up with many new results on heavy-ion collisions. 
Further updates of these facilities are planned or being proposed.
Compressed baryonic matter experiments at FAIR \cite{Friman:2011zz}
and NICA \cite{Kekelidze:2012zzb}, which will probe the QCD phase 
diagram in a high baryon density but relatively low temperature region, 
are a few years in the
future. Electron-ion collider (EIC) has been proposed to understand
the glue that binds us all \cite{Accardi:2012hwp}. So this exciting
field is going to remain very active for a decade at least.

\bigskip\bigskip

Many review articles have been cited throughout the text
above. Here are a few more published in the last 2-3 years
\cite{Muller:2012zq,Jacak:2012dx}.
See also these two talks given at the `2013 Nobel Symposium on LHC Physics'
for an overview of the status of this field:
\cite{Muller:2013dea,Schukraft:2013wba}.
% ----------------------------------------------------------------------------

\section*{Acknowledgements}

I sincerely thank Saumen Datta for a critical reading of the manuscript.

% ----------------------------------------------------------------------------

\end{document}